\documentclass[preprint,floatfix,nofootinbib,showpacs,preprintnumbers,amsmath,amssymb,prd,here, Here]{revtex4}

\usepackage{graphicx}
\usepackage{epsfig}
\usepackage{dcolumn}
\usepackage{bm}
\usepackage{subfigure}
\usepackage[usenames]{color}
\usepackage{color}

\newcommand{\be}{\begin{eqnarray}}
\newcommand{\ee}{\end{eqnarray}}
\newcommand{\sbe}{\begin{eqnarray*}}
\newcommand{\see}{\end{eqnarray*}}
\newcommand{\nn}{\nonumber}

\newcommand{\pslash}{\not\!p}
\newcommand{\qslash}{\not\!q}
\newcommand{\kslash}{\not\!k}

\usepackage[showlabels,sections,floats,textmath,displaymath,here]{preview}

\begin{document}

\title{
Strongly-Coupled  Unquenched QED$_4$ Propagators Using Schwinger-Dyson Equations 
}
\author{ Ay{\c s}e K{\i}z{\i}lers{\" u} }
\email{akiziler@physics.adelaide.edu.au}
\affiliation{
 Special Research Centre for the Subatomic Structure of Matter (CSSM)\\
   Department of Physics and Mathematical Physics, \\
   The School of Chemistry and Physics, \\
   Adelaide University, 5005, Australia }
   
%
%
\author{Tom Sizer}
 \email{tsizer@physics.adelaide.edu.au}
\author{Anthony G.\ Williams}
\email{anthony.williams@adelaide.edu.au}
\affiliation{
 Special Research Centre for the Subatomic Structure of Matter (CSSM)\\
   Department of Physics and Mathematical Physics, \\
   The School of Chemistry and Physics, \\
   Adelaide University, 5005, Australia
}


\date{\today}

\begin{abstract}
We study unquenched QED in four dimensions using renormalised Schwinger-Dyson equations and focus on the behaviour of the fermion and photon propagators.  For this purpose we use an improved K{\i}z{\i}lers{\"u}-Pennington (KP) vertex which respects gauge invariance, multiplicative renormalizability for the massless case, agrees with perturbation theory in the weak coupling regime and is free of kinematic singularities. We find that the KP vertex performs very well as expected specially in comparison with other vertex choices. We find that the Landau pole problem familiar from perturbative QED persists in the nonperturbative case with the renormalised inverse photon propagator having zero crossing.   
\end{abstract}

\pacs{Valid PACS appear here}


\maketitle

\section{\label{sec:Introduction}Introduction}

Studies of gauge field theories such as Quantum Electrodynamics (QED) and Quantum Chromodynamics (QCD) in the non-perturbative strong-coupling regime are of great interest as this is where the  
phenomena of confinement and Dynamical Symmetry Breaking occur. In order to explore the strong coupling region of gauge field theories one  needs non-perturbative tools like Lattice Gauge Theories (LGTs) in discrete space-time and Schwinger-Dyson Equations(SDE)\cite{Schwinger:1951ex,Schwinger:1951hq,Dyson:1949bp,Dyson:1949ha,Green:inteqn,Roberts:1994dr}, in the continuum. They are complementary techniques, each with their  own pros and cons.  While lattice has the strong appeal of being a first-principles approach, SDEs allow a much greater range of distance scales to be probed simultaneously. The SDEs are the field equations of a given Quantum Field theory, and as such, are a useful medium for studying non-perturbative Greens functions in the strong coupling regime over  a very wide range of momentum.  

The shortcoming of working with  these equations is that they form  an infinite tower of nested non-linear integral equations and hence need to be truncated so that they can be solved. Although Perturbation Theory is known as a consistent truncation scheme to these equations in the weak coupling regime, in order to understand the behaviour of field theories in the strong coupling regime one needs to treat the SDEs in such a way that they satisfy the greatest possible number of requirements including gauge invariance\cite{Delbourgo:1978bu,Salam:1963sa,Zumino:1959wt} , multiplicative renormalisability 
(MR)\cite{collins,zuber,Brown:1989hy,Curtis:1991fb,Kizilersu:2009kg}, consistency with perturbation theory in the weak-coupling regime and so on. The goal is to include as many theoretical constraints as possible so that the truncation preserves as much of the true physics of the theory as possible. In addition, in the longer term further constraints may emerge over a limited momentum-window from complementary lattice studies. 

The structure of the SDEs are such that  the 2-point Green's functions requires knowledge of 3-point Green's functions, the 3-point Green's functions in principle knows about $n$-point Green's functions and so on. However the most important question to answer for non-perturbative QED studies is
 ``{\it{what is the necessary and sufficient knowledge of the fermion-boson vertex in order to describe the complete and correct behaviour of the fermion and boson propagators?}}''. For more than four decades there have been many challenges to solving these equations using a variety of truncations and approximations, most of the efforts to date have concentrated on the fermion Schwinger-Dyson equation with an assumed form for the gauge boson propagator. 

The most rudimentary truncation scheme is called the Rainbow-Ladder approximation\cite{Fukuda:1976zb,Kondo:1990ig,Fomin:1984tv,Miransky:1984ef,Miransky:1986ib,Miransky:1986xp,Leung:1989hw,Kondo:1989my,Aoki:1989bu,Curtis:1993py,Hawes:1994ce}  which replaces the full (dressed) vertex with the {\it{Bare vertex}}  and full (dressed) photon propagator with the bare one. This is a quenched treatment since it ignores the fermion loops in  the photon propagator. One can therefore study this closed system for the fermion propagator, which consists of the two scalar functions (called the fermion wave-function renormalisation and the mass function). Use of this truncation makes it possible to perform some analytical calculations as well as the numerical ones. With the Rainbow-Ladder treatment, it was found that the fermion wave function renormalisation has a power-law behaviour in the asymptotic regions\cite{Brown:1989hy,Curtis:1991fb}, the dynamical mass also displays a power-law tail and the corresponding critical coupling above which the fermion mass dynamically generated is calculated to be $\pi/3$, 
\cite{Curtis:1993py,Miransky:1984ef,Miransky:1986ib,Miransky:1986xp}. However this truncation scheme does not satisfy the Ward-Green-Takahashi Identity (WGTI)\cite{Ward,Green:1953te,Takahashi:takahashi}, which is a relationship between the inverse full fermion propagators and the full fermion-photon vertex function. 
Ball and Chiu\cite{Ball:1980ay}, using the WGTI showed that the longitudinal part of the vertex can be uniquely specified (known as the Ball-Chiu vertex (BC)) whereas the transverse vertex  remained unconstrained.  On the other hand studies using both Bare and BC vertices yield gauge dependent critical coupling\cite{Curtis:1993py} while the critical coupling being a physical  quantity must be independent of  gauge parameter. 

Curtis and Pennington \cite{Curtis:1990zs,Curtis:1993py} presented an Ansatz for the transverse part of the  three-point Greens function  which is known as the CP vertex. 
Their argument was that multiplicative renormalisation of the propagator functions constrains the transverse part of the vertex, and therefore the transverse vertex can be built by making use of these constraints together with the other vertex requirements and help of perturbation theory in the weak coupling regime. 
The transverse part of the vertex consists of eight form factors however Curtis and Pennington only used one of them to construct their vertex, in other words with minimal contribution from the transverse vertex.  
Following this progress  Atkinson et.al. \cite{Atkinson:1993mz} showed that by including this minimal transverse vertex (CP) the gauge dependence of the critical coupling is reduced considerably.
Later on Burden and Roberts \cite{Burden:1993gy} used gauge covariance concepts to constrain the fermion-photon vertex. As an implementation and continuation of this work, Dong et.al.\cite{Dong:1994jr} wrote down a vertex Ansatz for massless quenched  QED which respects the Ward identity and makes the fermion propagator gauge covariant, yet their construction involved an unknown function.  Improvements to this study came from Bashir and Pennington\cite{Bashir:1994az, Bashir:1995qr}, who used the same arguments for massive QED and included more form factors and thereby constructed their transverse vertex in terms of two unknown functions. 

Although all these studies were very useful in many ways, namely in understanding the internal structure of SDE's, in understanding the role and importance of the vertex in the propagator functions, in learning about  the phase structure of the quenched theory and in building the technology in solving and dealing with these equations, all these studies were done using quenched approximations\cite{Dong:1994jr,Haeri:1990ty,Fischer:2004ym,Atkinson:1989fp,Roberts:1994dr,Atkinson:1993mz,Bloch:1994if,Hawes:1991,Hawes:1994ce,Hawes:1996ig,Hawes:1996mw,Alkofer:2000wg,Kizilersu:2000qd,Kizilersu:2001,Miransky:1986ib,Miransky:1984ef,Kondo:1991ms,Kondo:1991mq,Pisarski:1984dj,Pisarski:1991}. 
The few previous  unquenched studies\cite{Kondo:1990ig,Kondo:1990st,Bloch:1995dd}  either employs the one loop perturbative expansion of the photon propagator in solving fermion SDE propagator
or introducing some approximations such as simpler vertex, choosing specific gauge in  solving the coupled system of fermion and photon SDEs.  Studies with this minimal inclusion of the dressed photon propagator have  served as a valuable stepping stone, nevertheless in order to understand the behaviour of the strongly coupled fermion and photon system a more realistic unquenched fermion-photon vertex is needed. Recently such a vertex has become available through Kizilersu and Pennington(KP)\cite{Kizilersu:2009kg}, who constructed their fermion-photon vertex so as to ensure multiplicatively renormalisablity of the fermion and photon propagators, to respect gauge invariance and to be consistent with perturbation theory in the weak coupling regime. 

This paper provides a comprehensive study of strongly coupled  unquenched QED in 4-dimensions in general covariant gauges by employing the unquenched fermion-photon vertex of Kizilersu-Pennington\cite{Kizilersu:2009kg}. The results are contrasted together with the other commonly used vertices such as bare, Ball-Chiu, Curtis-Pennington for comparison. We will analyse this coupled physical system of SDEs thoroughly  by examining their unquenching effects, testing the vertices for their influence on the  behaviour of propagators.

This article is organised such that Sect.II introduces our notation, conventions and all the equations that they will be solved  later.
 In Sec.III we describe our approach and methodology for solving the Schwinger-Dyson equations for the propagator functions. 
 We specify the equations for fermion wave-function renormalisation, mass function and the photon wave-function renormalisation that need to be solved. 
 Section IV presents our numerical results and includes a discussion of these results. In Sec.V we conclude and outline future work.

\section{\label{sec:Notation} Schwinger-Dyson Equations Approach and its Conventions}

\begin{figure*}[htp]
\begin{center}
\includegraphics{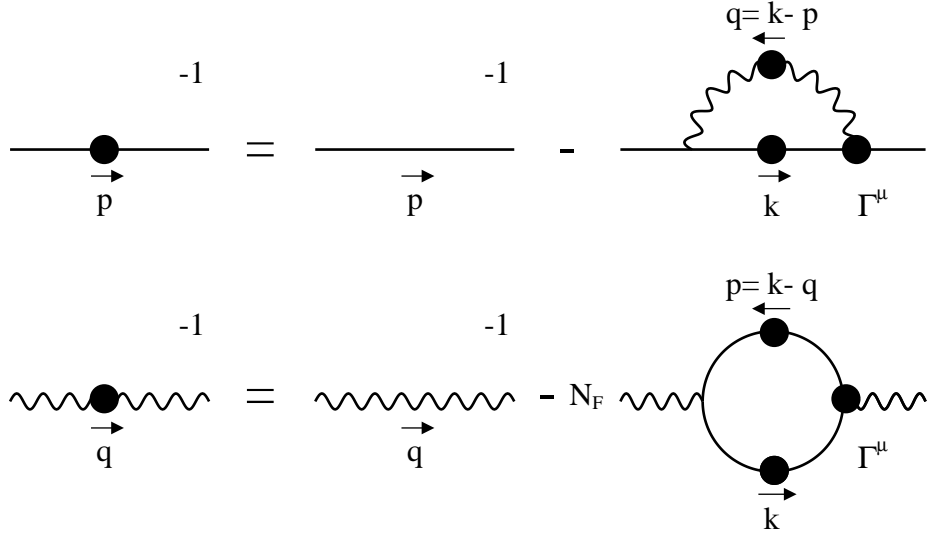}
\end{center}
\caption{The Schwinger-Dyson equations for the fermion and photon propagators in QED.} 
\label{fig:vertex}
\end{figure*}

The SDE equations for the 2-point Green's functions are shown diagrammatically in Fig.\ref{fig:vertex}. 
These diagrammatical equations display how the full (dressed) propagator functions on the LHS (solid and wavy lines with solid dotes) 
are connected to the bare and dressed fermion and boson propagators and to the dressed fermion-boson vertex function on the RHS.
\noindent
Making use of the Feynman rules for these graphs, the diagrammatic SDEs can be written down explicitly as a set of non-linear coupled integral equations for the inverse fermion propagator $S^{-1}$ and the inverse photon
propagator $\Delta_{\alpha\beta}^{-1}$ respectively,
\be S_F^{-1}(p) &=& {S^{0}}_F^{-1}(p) \> - \>i e^2 \int _0^{\infty}{d^4 k \over (2 \pi)^4}\> \Gamma^\alpha (p,k;q) \> S(k)\,\gamma^\beta \> \Delta_{\alpha \beta}(q)\,,
\label{eq:mainsdf} \\
&\equiv&  {S^{0}}^{-1}(p)\> - \>  \Sigma(p)\, ,\nonumber \\
\Delta_{\alpha \beta}^{-1}(q)  &=& {\Delta_{\alpha \beta}^{0}}^{-1}(q) \> + \>i e^2 N_F \> {\rm Tr} \int_0^{\infty} {d^4 k \over (2
\pi)^4}\> \Gamma_\alpha (p,k;q)\>S(k) \>\gamma_\beta \>  S(p)\>\,,
\label{eq:mainsdph} \\
&\equiv& {\Delta_{\alpha \beta}^{0}}^{-1}(q) \> +  \> \Pi_{\alpha \beta}(q)\,,\nonumber 
\ee
\noindent
where $e$ is a bare fermion charge, $\Gamma_\mu(p,k)$ is the full fermion-photon vertex, $q=k-p$ is the photon momentum, $\Sigma(p^2)$ is
the fermion self-energy and $\Pi_{\alpha \beta}(q^2)$ is the photon self-energy or photon polarization, $S^0$ and $\Delta_{\alpha\beta}^0$ are the tree level fermion and photon propagators respectively .

The ultimate goal is to solve the above coupled SDEs for the propagators, and to do this the necessary unknown functions we need are the fermion and the photon 
propagators and the fermion-photon vertex, which we will discuss next.
 %
%
\subsection{Fermion Propagator}
The full (dressed) fermion propagator can be defined in terms of two scalar functions $(F$ and $M)$ or equivalently $(A$ and $B)$ 
\begin{equation}\label{FermionPropagator}
    i\,S_F(p) = i\,\frac{F(p^2)}{\pslash - M(p^2)} = i\,\frac{1}{A(p^2)\pslash - B(p^2)}\,,
\end{equation}
where
\begin{equation}
    F(p^2)\equiv \frac{1}{A(p^2)}\,,\quad
    M(p^2) \equiv \frac{B(p^2)}{A(p^2)}\,.
\end{equation}
Here ${{F}(p^2)}$ is the fermion wave-function renormalization function
and ${{M}(p^2)}$ is the mass function.
\noindent
The bare, or tree-level form, of the fermion propagator where

\noindent
$(F(p^2)=1 \, {\rm{and}}\,  M(p^2)=m_0)$ is

\begin{equation}\label{BareFermionPropagator}
    i\,S^0(p) = i\,\frac{1}{\pslash - m_0} = i\,\frac{\pslash + m_0}{p^2 - m_0^2}\,,
\end{equation}
where $m_0$ is the bare mass.
%
%
\subsection{Photon Propagator}
The full photon propagator can be defined in terms of the scalar function, G 
\begin{equation}\label{PhotonPropagator}
    i\,\Delta_{\mu\nu}(q)
        = \frac{-i}{q^2}
            \left[
                G(q^2)\left(g_{\mu\nu}-\frac{q_{\mu} q_\nu}{q^2}\right)+
                \xi\frac{q_\mu q_\nu}{q^2}
            \right]\,.
\end{equation}
Here, $\xi$ is the covariant gauge parameter and $G(q^2)$ is the photon wave-function renormalization function which is related to the 
scalar self-energy part of the photon, $\Pi(q^2)$, by
\begin{equation}
\label{PhotonInvFieldStrength}
    G(q^2) = \frac{1}{1- \Pi(q^2)}\,.
\end{equation}
\noindent
The bare, or the tree-level form, of the photon propagator $(\rm{when}\, G(p^2)=1)$ is
\begin{equation}
    i\,\Delta_{\mu\nu}^0(q)
        = \frac{-i}{q^2}
            \left[
                \left(g_{\mu\nu}-\frac{q_{\mu} q_\nu}{q^2}\right)+
                \xi\frac{q_\mu q_\nu}{q^2}
            \right]\quad.
            \label{eq:invbareph}
            \end{equation}
\noindent
The inverse full photon propagator is
\begin{equation}\label{eq:InvPhotonPropagator}
    \left(i\,\Delta_{\mu\nu}\right)^{-1}(q)
        = i\,q^2
            \left[
                \frac{1}{G(q^2)}\left(g_{\mu\nu}-\frac{q_{\mu} q_\nu}{q^2}\right)+
                \frac{1}{\xi} \frac{q_\mu q_\nu}{q^2}
            \right]\,.
\end{equation}
One can also write the photon propagator in terms of its transverse and longitudinal parts~: 
\begin{equation}
    i\,\Delta_{\mu\nu}(q)=\frac{-i}{q^2}\left[G(q^2) \Delta_{\mu\nu}^T + \Delta_{\mu\nu}^L \right] \quad,
    \label{eq:photontl}
\end{equation}
where
\begin{equation}
    \Delta_{\mu\nu}^T(q) = g_{\mu\nu}-\frac{q_{\mu} q_\nu}{q^2}\,,\quad
    \Delta_{\mu\nu}^L(q) = \xi\frac{q_\mu q_\nu}{q^2} \quad .
\end{equation}
%
%
%
\subsection{The Full (Dressed) Fermion-Photon Vertex}

The complete QED vertex involves 12 independent vector structures 
which can be formed from the vectors $\gamma_\mu, k_\mu, p_\mu$ and the spin scalars $ 1, \not\!k, \not\!p$ and $\not\!k \not\!p $ 
\begin{eqnarray}
 \Gamma_F^\mu(p,k) =\sum_{i=1}^{12}\, f^i(p^2,k^2,q^2)\, V_i^{\mu}(p,k,q)\,,
 \label{eq:generalvertex}
 \end{eqnarray}
where $f^i$ are coefficient functions and $V_i^{\mu}$ are the spin structures. 
The full vertex may be split into transverse and longitudinal components,
\begin{equation}
    \Gamma_F^\mu(p,k) = \Gamma_T^\mu(p,k) + \Gamma_L^\mu(p,k)\,,
\end{equation}
with 
\begin{equation}
    q_\mu \, \Gamma_T^\mu(p,k) = 0\,.
    \label{eq:transverseWGT}
\end{equation}
In gauge theories the full vertex satisfies  
the Ward-Green-Takahashi identity(WGTI) \cite{Ward,Takahashi:takahashi,Green:1953te} which is a relation between the longitudinal part of the dressed vertex function through  Eq.(\ref{eq:transverseWGT}) and the inverse fermion propagator
\begin{equation}
\label{eq:vertexWTI}
    q_\mu \, \Gamma_F^\mu(p,k) = q_\mu \, \Gamma_L^\mu(p,k) = S^{-1}(k) - S^{-1}(p) \quad {\rm{with}} \quad (q = k - p)\,,
\end{equation}  
and the Ward identity, which is the nonsingular $q \longrightarrow 0$ i.e. $k \longrightarrow p$ limit of WGTI~: 
\be
\Gamma_\mu(p,p) = \lim_{k \longrightarrow p}\, \Gamma_\mu(k,p) = \frac{\partial S_F^{-1}(p)}{\partial p_\mu}\quad\,\,
{\rm {with}} \quad\,\,
\Gamma^\mu_T(p,p) = 0\,.
\label{eq:wardidentity}
\ee
Therefore both the Ward-Green-Takahashi and Ward identities ensure that the full vertex and the longitudinal vertex are free of {\it{ kinematic singularities}}, and in return the transverse vertex should be also free of kinematic singularities. 

\subsubsection{\underline{The Longitudinal Vertex}}

The WGTI, Eq.(\ref{eq:vertexWTI}), is a statement about the longitudinal part of the vertex   
and it does not constrain the transverse part except that in the limit ${k \longrightarrow p}$ the transverse vertex vanishes. Implementing this idea of 
longitudinal vertex being free of kinematic singularities led Ball and Chiu \cite{Ball:1980ay} to {\it{uniquely}}  decompose the longitudinal vertex (the Ball-Chiu vertex) in terms of  some specific linear combination of some spin amplitidues, $V_i$, in Eq.(\ref{eq:generalvertex}), and their coefficient functions, $f_i$, as~:
\begin{equation}
    \Gamma_L^\mu(p,k) =
        \lambda_1(p^2,k^2) \gamma^\mu +
        \lambda_2(p^2,k^2) (\kslash + \pslash) (k+p)^\mu +
        \lambda_3(p^2,k^2) (k+p)^\mu +
        \lambda_4(p^2,k^2)  (k^{\nu}+p^{\nu})\sigma^{\mu\nu}\,,
        \label{eq:longitudinal}
\end{equation}
where the longitudinal form factors $\lambda_i$   in Minkowski space are
\be
    \lambda^M_1(p^2,k^2) &=&
                   \frac{1}{2}          \left[A(k^2) + A(p^2) \right]\,, \\[2mm]
    \lambda^M_2(p^2,k^2) &=&
                   \frac{1}{2(k^2-p^2)} \left[A(k^2) - A(p^2) \right] \,,\\[2mm]
    \lambda^M_3(p^2,k^2) &=&
                   \frac{-1}{k^2-p^2}   \left[M(k^2)\,A(k^2) - M(p^2)\,A(p^2) \right]\,, \\[1mm]
    \lambda^M_4(p^2,k^2) &=& 0\,\,.
\ee
\noindent
These longitudinal form factors were determined by Ball and Chiu in terms of fermion wave function renormalization and mass function, and hence 4 of the 12 tensor structures in the full vertex appear in this BC vertex construction. Furthermore,  all of the singularities (IR not kinematic ones) in the full vertex are expected to be encapsulated in the longitudinal vertex. Their conjecture is supported by the one-loop perturbative calculation\cite{Ball:1980ay,Kizilersu:1995iz} of the fermion-photon vertex indicating no such kinematic singularities.  
\subsubsection{\underline{The Transverse Vertex}}

The remaining 8 vector structures are used to construct the transverse part of the vertex, it may be written in generality as (with $q = k - p$)
\begin{equation}
    \Gamma_T^\mu(p,k) = \sum_{i=1}^8 \tau_i(p^2,k^2,q^2) \, T_i^\mu(p,k)\,,
    \label{eq:transverse}
\end{equation}

\noindent
where the form factors, $\tau_i$, are unknown scalar functions and $T_i^\mu$'s

\be
    T_1^\mu(p,k) &=& p^\mu \left(k\cdot q\right) - k^\mu \left(p\cdot q\right) \,,\nn\\
    T_2^\mu(p,k) &=& \left[ p^\mu \left(k\cdot q\right) - k^\mu \left(p\cdot q\right) \right] \,
                     \left(\kslash + \pslash\right) \,,\nn\\
    T_3^\mu(p,k) &=& q^2 \gamma^\mu - q^\mu \qslash \,,\nn\\
    T_4^\mu(p,k) &=& q^2 \left[\gamma^\mu \left(\kslash + \pslash\right) - \left(p+k\right)^\mu\right] +
                     2 \left(p-k\right)^\mu k^\lambda p^\nu \sigma_{\lambda\nu} \,,\nn\\
    T_5^\mu(p,k) &=& q_\nu \sigma^{\nu\mu} \,,\nn\\
    T_6^\mu(p,k) &=& \gamma^\mu \left(p^2 - k^2\right) + \left(p+k\right)^\mu \qslash \,,\nn\\
    T_7^\mu(p,k) &=& \frac{1}{2} \left(p^2-k^2\right) \left[\gamma^\mu \left(\kslash + \pslash\right) - \left(p+k\right)^\mu\right] +
                     \left(k+p\right)^\mu k^\lambda p^\nu \sigma_{\lambda\nu} \,,\nn\\
    T_8^\mu(p,k) &=& -\gamma^\mu k^\nu p^\lambda \sigma_{\nu\lambda} + k^\mu \pslash - p^\mu \kslash \,,
    \label{eq:Ts}
\ee
are the transverse basis vectors 
which were previously defined by Ball and Chiu in \cite{Ball:1980ay} as linear combinations of $V_i^\mu$ in Eq.(\ref{eq:generalvertex}) in such a way that the transverse vertex is NOT contributing to the WGTI namely satisfying $ (q_\mu \, \Gamma_T^\mu(p,k) = 0)$ and it vanishes in the limit 
$ k \longrightarrow p$ i.e. satisfying $ \Gamma^\mu_T(p,p) = 0 $. Their form were also guided by perturbation theory to avoid kinematic singularities in the individual form factors as well since these singularities do not arise in one-loop perturbative calculations of transverse vertex\cite{Ball:1980ay,Kizilersu:1995iz}. The higher-order perturbative calculations are not expected to introduce any such singularities since WGT and Ward identities are non-perturbative expressions and as such they have to be respected at all orders by the same mechanism as the lowest-order terms. 

Although the full transverse vertex is expected to be free of a kinematic singularities the individual form factors do not have to be however the choice of the basis tensors by Ball and Chiu possess this feature and as consequently their one-loop  perturbative form factors in the Feynman gauge $\xi=1$ do not exhibit any such singularities. On the other hand the complete calculations of one-loop fermion-photon vertex in general covariant gauge given by K{\i}z{\i}lers{\" u} et.al.\cite{Kizilersu:1995iz} exhibited that for this choice of basis tensors there are singularities in that $\tau_4$ and $\tau_7$  each have a singularity separately which cancels in the full transverse vertex. They proposed  alternative new basis tensors, $T_4^\mu$ and $T_7^\mu$, in which these singularities do not appear.  
Note that $T_{1,2,3,4}$ are symmetric under $k \leftrightarrow p$, while $T_{5,6,7,8}$ are antisymmetric under the same transformation.

Knowledge of the fermion-boson vertex is essential to solving the coupled Schwinger-Dyson equations for the propagator functions, Eqs.(\ref{eq:mainsdf} \& \ref{eq:mainsdph}). Since the 1950s there have been many SDEs studies, which employed various vertices and these are summarized below.   

\subsubsection{\underline{Vertices Under Consideration}}

\begin{itemize}
\item{ \underline{\bf{Bare Vertex}}}

This is the minimal vertex contribution within the full vertex construction and is the first order contribution in perturbation theory:
\begin{eqnarray}
  \Gamma_{\rm F}^\mu = \Gamma_{\rm Bare}^\mu &=& \gamma^\mu\,.
\end{eqnarray}
It is clearly inadequate as it doesn't satisfy the WGTI except in the massless quenched approximation in the Landau gauge $(\xi=0)$, nor does it satisfy multiplicative renormalizability (MR). However, in the Landau gauge at least, it reproduces qualitatively the features of quenched (where the fermion loops in photon propagator are neglected, i.e. the photon propagator is treated as the bare one) QED, in that the spinor part of the WGTI is satisfied. There have been many studies employing Bare vertex and some are \cite{Fukuda:1976zb,Kondo:1990ig,Fomin:1984tv,Miransky:1984ef,Miransky:1986ib,Miransky:1986xp,Leung:1989hw,Kondo:1989my,Aoki:1989bu,Curtis:1993py,Hawes:1994ce}.


\item{\underline{ \bf{Ball-Chiu Vertex (BC)}}}

\noindent
Strictly speaking, the Ball-Chiu (BC) \cite{Ball:1980ay} vertex is the longitudinal part of the fermion-photon vertex, Eq.~(\ref{eq:longitudinal}) with no  transverse contribution:
\begin{eqnarray}
   \Gamma_{\rm F}^\mu = \Gamma_{\rm BC}^\mu = \Gamma_{\rm L}^\mu 
    &=&  \frac{\gamma^\mu}{2}\, \left[A(k^2) + A(p^2) \right] 
    +  \frac{(\kslash + \pslash) (k+p)^\mu}{2(k^2-p^2)} \left[A(k^2) - A(p^2) \right]\nonumber\\
        &-& \frac{(k+p)^\mu}{k^2-p^2}   \left[M(k^2)\,A(k^2) - M(p^2)\,A(p^2) \right] \,,
\end{eqnarray}
Although this vertex satisfies the WGT and Ward identities and hence is free of any kinematic singularity, it does not satisfy Multiplicative Renormalizability.


\item{\underline{ \bf{Curtis-Pennington Vertex (CP)}}}

\noindent
The Curtis-Pennington vertex\cite{Curtis:1990zs} is the BC longitudinal vertex with a minimal transverse part with only one non-zero form factor:
\begin{eqnarray}
   \Gamma_{\rm F}^\mu = \Gamma_{\rm CP}^\mu &=&  \Gamma_L^\mu + T_6^\mu \times \tau_6\,,
\end{eqnarray}
where
\begin{equation}
   \tau_6^M = -\frac{1}{2\,d(k^2,p^2)} \left(A(k^2) - A(p^2)\right)\,,
\end{equation}
and ``$M$'' denotes Minkowski space and
\begin{equation}
    d(k^2,p^2) = \frac{\left(k^2-p^2\right)^2 + \left[M^2(k^2) + M^2(p^2)\right]^2}{k^2 + p^2}\,.
\end{equation}
This vertex is  designed to be multiplicative renormalizable and it is proven to be very successful in many non-perturbative  quenched QED studies, it has a dynamical problem when used in unquenched studies which we will discuss in Sec.\ref{sec:KP vertex comparison} in detail.


\item{\underline{ \bf{Modified Curtis-Pennington Vertex (Mod. CP)}}}

\noindent
Because of the undesirable feature of the CP vertex  in unquenched studies noted above, a modified version is used in an ad-hoc (hybrid) fashion. This hybrid vertex consists of the Curtis-Pennington  construction for the fermion Schwinger-Dyson Equations and the Ball-Chiu construction for the photon SDE which is also studied in\cite{Bloch:1995dd}.
\begin{eqnarray}
   \Gamma_{\rm F}^\mu &=& \Gamma_{\rm CP}^\mu =  \Gamma_L^\mu + T_6^\mu \times \tau_6\,, \qquad {\rm for\,\, fermion\,\, SDE} \nonumber\\
    \Gamma_{\rm F}^\mu &=& \Gamma_{\rm BC}^\mu =  \Gamma_L^\mu\, \qquad \qquad\qquad\quad {\rm for \,\, photon\,\, SDE}\,.
    \label{eq:modCP}
\end{eqnarray}
%


\item{\underline{\bf{K{\i}z{\i}lers{\"u}-Pennington Vertex (KP)}}}

This is a newly proposed vertex \cite{Kizilersu:2009kg} was designed for unquenched studies: its form factors carry both fermion and photon momenta dependence. It satisfies both fermion and photon SDEs to all orders in leading logarithms and is multiplicatively renormalizable by construction in  the massless case and respects the WGT and Ward identities. The KP vertex studies  concluded that there is more than one vertex construction that satisfies the unique photon limit 
$\protect{k^2 \simeq p^2 \gg q^2}$ and all necessary constraints but that they differ from each other only beyond the leading logarithmic order. Two such constructions  mentioned in the original paper\cite{Kizilersu:2009kg} and studied numerically here are given below~:
\begin{eqnarray}
  \Gamma_{\rm F}^\mu  =  \Gamma_{\rm KP}^\mu &=&  \Gamma_{\rm BC}^\mu + T_2^\mu \, \tau_2 + T_3^\mu \, \tau_3 + T_6^\mu \, \tau_6 + T_8^\mu \, \tau_8\,,
\end{eqnarray}
where\\
{\bf{TYPE 2}}
 \be
\tau_2^E(p^2,k^2,q^2) =& - &\frac{4}{3}\,\frac{1}{(k^4-p^4)}\,\left(A(k^2)\,-\,A(p^2)\right)\nn\\[3mm]
                                   &-& \frac{1}{3} \,\frac{1}{(k^2+p^2)^2}\,  \left(A(k^2) + A(p^2) \right)\,\ln \left[\left(\frac{A(k^2)\,A(p^2)}{A(q^2)^2}\right)\right]\,,  \nn\\ [3mm]
\tau_3^E(p^2,k^2,q^2)=&-& \frac{5}{12}\, \frac{1}{(k^2-p^2)}    \; \left(A(k^2)\,-\,A(p^2)\right) \nn\\ [3mm]
                                   &-&\frac{1}{6}\,\frac{1}{(k^2+p^2)}\;  \left(A(k^2) + A(p^2) \right)\,   \ln \left[\left(\frac{A(k^2)\,A(p^2)}{A(q^2)^2}\right)\right] \,, \nn\\[3mm]
\tau_6^E(p^2,k^2,q^2) = && \frac{1}{4} \;\frac{1}{(k^2+p^2)}\,   \left(A(k^2)\,-\,A(p^2)\right)\,, \nn\\[3mm]
\tau_8^E(p^2,k^2,q^2) = &&0\,,
\label{eq:finaltaus3}
\ee

{\bf{TYPE 3}}
\be
\tau_2^E(p^2,k^2,q^2) =& - &\frac{4}{3}\,\frac{1}{(k^4-p^4)}\,\left(A(k^2)\,-\,A(p^2)\right)\nn\\[3mm]
                                   &- & \frac{2}{3} \,\frac{1}{(k^2+p^2)^2}\,  \left(A(k^2) + A(p^2) \right)\,\ln\left[\frac{1}{2}\left(\frac{A(k^2)}{A(q^2)}+\frac{A(p^2)}{ A(q^2)}\right)\right]\,,  \nn\\ [3mm]
\tau_3^E(p^2,k^2,q^2)=&-& \frac{5}{12}\, \frac{1}{(k^2-p^2)}    \; \left( A(k^2)\,-\,A(p^2)\right) \nn\\ [3mm]
                                   &-&\frac{1}{3}\,\frac{1}{(k^2+p^2)}\;  \left(A(k^2) + A(p^2) \right)\,   \ln \left[\frac{1}{2}\left(\frac{A(k^2)}{A(q^2)}+\frac{A(p^2)}{ A(q^2)}\right)\right] \,, \nn\\[3mm]
\tau_6^E(p^2,k^2,q^2) =& &\frac{1}{4} \;\frac{1}{(k^2+p^2)}\,   \left(A(k^2)\,-\,A(p^2)\right)\,, \nn\\[3mm]
\tau_8^E(p^2,k^2,q^2) = &&0\,,
\label{eq:finaltaus2}
\ee
 \noindent
where ``$E$'' denotes the Euclidean space form and these two types of vertices only differs in their arguments of $\ln$'s and in the coefficient factors.  Although we will be comparing above TYPE 2 and TYPE 3 vertices in Sec.(\ref{sec:KP vertex comparison}),  through out of our numerical studies in Sec.~(\ref{sec:Numerical solutions})  we will be using the TYPE 2 KP vertex.
 
\end{itemize}

Later in the paper, in Sec.(\ref{sec:Numerical solutions}) we will make use of these vertex ansatze listed above  to analyse their performance in unquenched propagator studies.

\section{\label{sec:Formulation} Regularization-Independent Method for Unquenched Fermion and Photon Propagators}

In Quantum Field theories the self energies involve divergent integrals Eqs.(\ref{eq:mainsdf},\ref{eq:mainsdph}) therefore the two step procedure of  regularization and renormalization are unavoidable.   One can employ regularization schemes such as ``Dimensional Regularization'' or  ``Cut-off Regularizaton'' 
which both have their own pros and cons in SDE studies. For instance, while Dimensional Regularisation respects the {\it {translational invariance}} its numerical implementation in SDE studies is challenging and it breaks chiral symmetry for all values of coupling, $\alpha$, 
in quenched QED\cite{Schreiber:1998ht} until the $\epsilon \longrightarrow 0$ limit is taken. On the other hand the Cut-off regularization does not respect the translation invariance but makes the numerical studies tractable\cite{Kizilersu:2000qd,Kizilersu:2001}. However one must use it with care, Fig.(4) in Ref.(\cite{Kizilersu:2001}) shows how the correct treatment of cut-off regularisation gives excellent agreement with the Dimensional Regularisation method. The second step in this procedure is to remove this regulator by renormalising the theory at the physical scale $\mu$.
 
In these unquenched studies of 4-dimensional QED  we will necessarily need to work with the renormalized quantities in SDEs in order to study regularisation-independent quantities and hence we next establish the renormalization procedure.

\subsection{\label{sec:renormalization} Renormalization:}

Our renormalization treatment is the standard one as we relate the regularized unrenormalized quantities to the  renormalized ones in the following multiplicative way~:
\be
{S}(p^2\,;\mu^2) \, &=& \, Z_2^{-1}(\mu^2,\Lambda^2) \, S^{\rm Bare}(p^2\,;\Lambda^2)\,, \\
{\Delta}_{\nu\sigma}(p^2\,;\mu^2) \, &=& \, Z_3^{-1}(\mu^2,\Lambda^2) \, \Delta_{\nu\sigma}^{\rm Bare}(p^2\,;\Lambda^2)\,, \\
{\Gamma}_\nu(p^2,k^2\,;\mu^2) \, &=& \,Z_1(\mu^2,\Lambda^2) \, \Gamma_\nu^{\rm Bare}(p^2,k^2\,;\Lambda^2)\,, \\
\xi &=& Z_3^{-1}(\mu^2,\Lambda^2)\, \xi^{\rm Bare}\,,\\
\alpha &=& Z_3(\mu^2,\Lambda^2)\,  \alpha^{\rm Bare} \,, \quad\quad (Z_1=Z_2) \,,
\ee
where $Z_1$, $Z_2$ and $Z_3$ are the renormalization functions for the vertex, fermion and photon respectively, $\alpha=e^2/(4\pi)$ is the coupling,
$\mu^2$ is the renormalization point and $\Lambda^2$ is our regularization parameter, which is a UV cut-off. Note that renormalized quantities have an implicit dependence on $\mu$, which we do not show for notational convenience and where we will work at sufficiently large $\Lambda$ such that the residual regularization parameter dependence on the renormalized quantities is negligible. 
The renormalization conditions that we use here at $p^2=\mu^2$ are~:
\be
F(\mu^2,\mu^2)&=& 1 \,, \nonumber \\
G(\mu^2,\mu^2)&=& 1  \,,\nonumber \\
M(\mu^2)&=& m_\mu\,.
\label{eq:renormcond}
\ee

\noindent
Making use of the above renormalisation relations the renormalized inverse fermion and photon propagators, Eqs.(\ref{eq:mainsdf},\ref{eq:mainsdph}) are:

\be
S_F^{-1}(p\,;\mu)&=&Z_2(\mu)\, S_0^{-1}(p)-Z_1(\mu)\,{\overline{\Sigma}}(p)\,,
\label{eq:mainrensdf} \\
\Delta_{\alpha \beta}^{-1}(q\,;\mu)  &=&  Z_3(\mu) \,{\Delta_{\alpha \beta}^{0}}^{-1}(q) \> + \>  Z_1(\mu) \> \overline{ \Pi}_{\alpha\beta}(q)\,,
\label{eq:mainrensdph}
\ee
where $Z_1=Z_2$ from WGTI and for notational convenience we will suppress the regularization dependence from now on leaving it implicit and writing $Z_1(\mu^2,\Lambda^2)$ as 
$Z_1(\mu)$, likewise the renormalised quantities ${\overline{\Sigma}}(p,\mu)$ and ${\overline{\Pi}}_{\alpha\beta}(p,\mu)$ as ${\overline{\Sigma}}(p)$ and ${\overline{\Pi}}_{\alpha\beta}(p)$, etc. 
%
\subsection{\label{sec:Fermion equation} Regularization-Independent Formulation of the Full Fermion Schwinger-Dyson Equation}
 \begin{figure}[htbp]
 \begin{center}
 \includegraphics{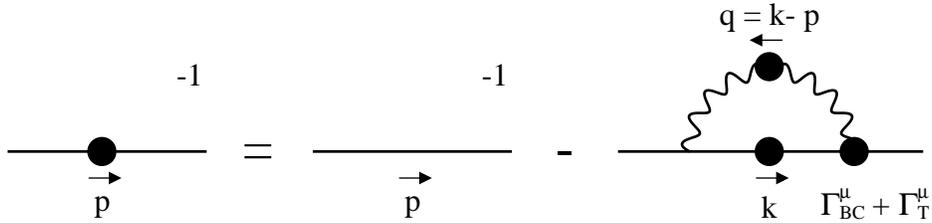}
 \caption{Fermion Schwinger-Dyson Equation.}
 \label{default}
 \end{center}
 \end{figure}
The fermion self-energy in Eq.(\ref{eq:mainrensdf}) can be decomposed into Dirac and scalar terms,
\noindent
$\protect{\overline{\Sigma}(p)= \overline{\Sigma}_d(p)\,{\not\!p}+{\overline{\Sigma}}_s(p)}$ which is obtained from $\Sigma(p)$ by

\begin{equation}
{\overline{\Sigma}}_d(p^2)=\frac{1}{4}\,{\rm Tr} \left(\overline{\Sigma}(p)\,\frac{\not\! p}{p^2}\right),  \qquad
{\overline{\Sigma}}_s(p^2)=\frac{1}{4}\,{\rm Tr} \left(\overline{\Sigma}(p)\cdot \mathbf{1}\right)\,.
\label{eq:sigmas}
\end{equation}

\noindent
Multiplying Eq.(\ref{eq:mainrensdf}) by $\not\!p$ and $\mathbf{1}$ respectively yields two separate
equations for the inverse fermion wave-function renormalization and the mass function~:
\be
F^{-1}(\mu^2;p^2) &=&  Z_2(\mu) \>-\> Z_2(\mu) \>\bar \Sigma_d(p^2)\,,
\label{eq:zeq}\\
B(p^2)=\,M(p^2)\> F^{-1}(\mu^2;p^2)&=&  Z_2(\mu)\> m_0\> +\> Z_2(\mu)\>\overline{ \Sigma}_s(p^2)\;\;\;.
\label{eq:meq}
\ee

\noindent
Evaluating Eqs.(\ref{eq:zeq},\ref{eq:meq}) at the renormalization point, $p^2=\mu^2$,
and forming an appropriate difference one can eliminate the divergent constants 
$Z_1$ and $Z_2$ to obtain the renormalized quantities
\be
\fbox{$\begin{array}{lll}
\displaystyle
 F(\mu^2;p^2) &=&  \displaystyle 1 \> + \> F(\mu^2;p^2) \>\overline{ \Sigma}_d(p^2) \> - \>
\overline{ \Sigma}_d(\mu^2)\,,
\label{eq:subwavefunc} \\
\displaystyle
M(p^2) & = & \displaystyle m_\mu \> + \> \left [M(p^2) \overline{ \Sigma}_d(p^2) \> + \>\overline{ \Sigma}_s(p^2)\right] \> -
\> \left[m_\mu \overline{ \Sigma}_d(\mu^2) \> + \> \overline{ \Sigma}_s(\mu^2)\right]\,,
\end{array}$}
\label{eq:submass}
\ee
\noindent
where the renormalization conditions Eq.(\ref{eq:renormcond}) have been realised. The left hand side of the
above equations being finite implies that the right hand side must also be finite, even though the individual 
$\overline{ \Sigma}_s$ and $\overline{ \Sigma}_d$
terms on the RHS may diverge separately as $\Lambda \longrightarrow \infty$. The details of the regularisation independent method can be found in \cite{Kizilersu:2001,Kizilersu:2002}.
 The equations in Eq.(\ref{eq:submass}) are the two main equations that we will be using for the fermion propagator in our analsis.

\noindent
{\bf{\underline{F Equation- The Fermion Wave-function Renormalisation:}}}

The fermion wave-function renormalisation, $F$, is defined in terms of the Dirac part of the self energy in Eq.(\ref{eq:submass}). 
Therefore starting with the Dirac part of the fermion self-energy, ${\overline{\Sigma}}_d$, in Eq.(\ref{eq:sigmas}), we write it explicitly 
in terms of dressed renormalized fermion-photon vertex, renormalized dressed fermion and photon propagators as
\be
\overline{\Sigma}_d(p^2)=\frac{i \alpha \pi}{p^2}\> {\rm Tr} {\not\!p} \int_M \frac{d^4k}{(2 \pi)^4} \,
\Gamma^\mu(p,k;\mu)\, S_F(k;\mu)\, \gamma^\nu \, \Delta_{\mu\nu}(q;\mu)\,.
\ee

\noindent
where $\alpha \equiv \alpha(\mu)$ is the running coupling defined at the renormalisation point. We employed the WGTI for the longitudinal part of the photon 
propagator, Eq.(\ref{eq:photontl}), and removed the odd integral  $\int d^4k \not\! q/q^4$ which would be zero under the translational invariant regularization scheme.  
After performing the trace algebra and moving from Minkowski space to Euclidean space using the Wick rotation we have~:
\be
\overline{\Sigma}_d(p^2)&=&  \frac{\alpha}{4\pi^3}\, \int_E d^4k\,\frac{1}{p^2} \frac{1}{q^2}\,\frac{F(k^2)}{
\left[k^2+M^2(k^2)\right]}\,\Big\{ \mathcal{I}^{L}_ {{\overline{\Sigma}}_d}+  \mathcal{I}^T_{{\overline{\Sigma}}_d}
\Big\} \,,\label{eqn:dirac_self_energy}
\ee
where $\mathcal{I}^{L}_{{\overline{\Sigma}}_d}$ and $\mathcal{I}^{T}_{{\overline{\Sigma}}_d}$ are the integrands
related to the longitudinal and transverse components of the fermion-photon vertex of the Dirac
part of the self-energy, ${\overline{\Sigma}}_d(p)$ respectively and they can be written as~:
%
%
\be
\label{eqn:sigmad_L} 
\mathcal{I}^{L}_{{\overline{\Sigma}}_d} &=&
                    - \frac{\xi}{q^2} A(p^2)\, \left\{  p^2\, k \cdot q +M(k^2)M(p^2)\,p \cdot q \right\}
\nonumber\\
                   &&+  G(q^2) \Bigg\{\,\,\,
                            \frac{1}{2}\,    \left[A(k^2)+A(p^2)\,\right]\, \frac{1}{q^2} \left[-2\Delta^2-3q^2 k \cdot p  \right]  \nonumber\\
               &&   \hspace{16mm} +\frac{1}{2\,(k^2-p^2)} \left[A(k^2)-A(p^2)\right]\frac{1}{q^2} \left[ -2 \Delta^2 (k^2+p^2) \right] \nonumber\\
                  && \hspace{16mm} + \frac{1}{(k^2-p^2)}    \left[M(k^2)^2\,A(k^2)-M(p^2)\,M(k^2)\,A(p^2)\right]\, \frac{1}{q^2} \left[ - 2\Delta^2\right]
                              \Bigg\}\,, \\   [3mm]
\mathcal{I}^{T}_{{\overline{\Sigma}}_d }&=&
   G(q^2) \Bigg\{\,\,\,\,\,\,\,
        \tau_1^E(p^2,k^2,q^2) M(k^2) \left[\Delta^2\right]  \nonumber\\
  &&  \hspace{15mm} + \tau_2^E(p^2,k^2,q^2) \left[-\Delta^2 (k^2+p^2) \right] \nonumber\\
  &&   \hspace{15mm} + \tau_3^E(p^2,k^2,q^2) \left[2\Delta^2 +3q^2 k \cdot p \right] \nonumber\\
  &&  \hspace{15mm}  + \tau_4^E(p^2,k^2,q^2) M(k^2)\left[2\Delta^2 +3q^2 (k \cdot p+p^2) \right] \nonumber\\
  &&   \hspace{15mm} + \tau_5^E(p^2,k^2,q^2) M(k^2)\left[3\,p \cdot q \right] \nonumber\\
  &&   \hspace{15mm} + \tau_6^E(p^2,k^2,q^2) \left[3 \, k \cdot p \,(p^2-k^2) \right] \nonumber\\
  &&   \hspace{15mm} + \tau_7^E(p^2,k^2,q^2) M(k^2)\left[-\Delta^2 -\frac{3}{2} (k^2-p^2) (p^2+ k \cdot p) \right] \nonumber\\
  &&  \hspace{15mm}  + \tau_8^E(p^2,k^2,q^2) M(k^2)\left[-2 \Delta^2 \right]\quad \Bigg\}\,.
 \label{eqn:sigmad_T}
\ee
Note that $\tau$'s are in Euclidean space and $\Delta^2=(k \cdot p)^2-k^2p^2$.

\noindent
{\bf{\underline{M Equation - Mass Function:}}}

In a similar way to fermion self energy, the mass function in Eq.(\ref{eq:submass}) is given by both Dirac and the scalar part of the self energy. Hence, 
the scalar part of the fermion self-energy, ${\overline{\Sigma}}_s$, in Eq.(\ref{eq:sigmas}) can be dealt in a similar way to ${\overline{\Sigma}}_d$
\be
\overline{\Sigma}_s(p)=i \alpha\pi\> {\rm Tr} \int_M \frac{d^4k}{(2 \pi)^4} \, \Gamma^\mu(p,k;\mu)\,
S_F(k;\mu)\, \gamma^\nu \, \Delta_{\mu\nu}(q;\mu) \,,
\ee

\noindent
and in Euclidean space it is

\be 
\overline{\Sigma}_s(p)&=&  \frac{\alpha}{4\pi^3}\, \int_E d^4k\ \frac{1}{q^2}\,\frac{F(k^2)}{
\left[k^2+M^2(k^2)\right]}\,\Big\{ \mathcal{I}^{L}_ {{\overline{\Sigma}}_s}+  \mathcal{I}^T_{{\overline{\Sigma}}_s}
\Big\} \,,
\label{eqn:scalar_self_energy} 
\ee

\noindent
where again $\mathcal{I}^{L}_{\overline{\Sigma}_s} $ and $\mathcal{I}^{T}_{\overline{\Sigma}_s} $ are the
integrands related to the longitudinal and transverse part of the fermion-photon vertex of the scalar part of the self energy, ${\overline{\Sigma}}_s(p)$, respectively and they are~:

\be
\mathcal{I}^{L}_{{\overline{\Sigma}}_s }
                               &=&  \frac{\xi}{q^2}\,\frac{1}{F(p^2)}\, \left[k \cdot q\, M(p^2)-p \cdot q M(k^2) \right]
\nonumber\\
&+&  G(q^2) \Bigg\{
                 \frac{1}{2}\, \left[ A(k^2)+A(p^2) \right] M(k^2) \left[3  \right]  \nonumber\\
&&     \hspace{15mm}     + \frac{1}{2(k^2-p^2)}\,\left[ A(k^2) - A(p^2) \right] M(k^2) \left[\frac{-4 \Delta^2}{q^2} \right] \nonumber\\
&&     \hspace{15mm}       + \frac{1}{(k^2-p^2)}\,
\left[ M(k^2)\,A(k^2) - M(p^2)\,A(p^2) \right]\,\left[\frac{2\Delta^2}{q^2}\right]
\Bigg\}\,,
\label{eq:sigmas_L} \\[3mm]
\mathcal{I}^{T}_{{\overline{\Sigma}}_s }
  &=& G(q^2) \Bigg\{\,\,\,\,\,\,\,
                     \tau_1^E(p^2,k^2,q^2) \left[-\Delta^2\right]  \nonumber\\
  && \hspace{15mm}       + \tau_2^E(p^2,k^2,q^2) \left[-2 \Delta^2 \right] M(k^2) \nonumber\\
  &&  \hspace{15mm}      + \tau_3^E(p^2,k^2,q^2) \left[-3 q^2 \right] M(k^2) \nonumber\\
  && \hspace{15mm}       + \tau_4^E(p^2,k^2,q^2) \left[2\Delta^2 + 3 q^2 (k \cdot p+k^2) \right] \nonumber\\
  && \hspace{15mm}       + \tau_5^E(p^2,k^2,q^2) \left[3 k \cdot q \right] \nonumber\\
  && \hspace{15mm}       + \tau_6^E(p^2,k^2,q^2) \left[-3 (p^2-k^2) \right]M(k^2) \nonumber\\
  && \hspace{15mm}       + \tau_7^E(p^2,k^2,q^2) \left[\Delta^2 +\frac{3}{2} (p^2-k^2) (k^2+ k \cdot p) \right] \nonumber\\
  && \hspace{15mm}       + \tau_8^E(p^2,k^2,q^2) \left[\,\,\,0 \,\,\,\right] \Bigg\}\,.
 \label{eq:sigmas_T}
 \ee

\subsection{\label{sec:Photon equation} Regularization-Independent Formulation for the Full Photon Schwinger-Dyson Equation}

 \begin{figure}[htbp]
 \begin{center}
 \includegraphics[width=14cm]{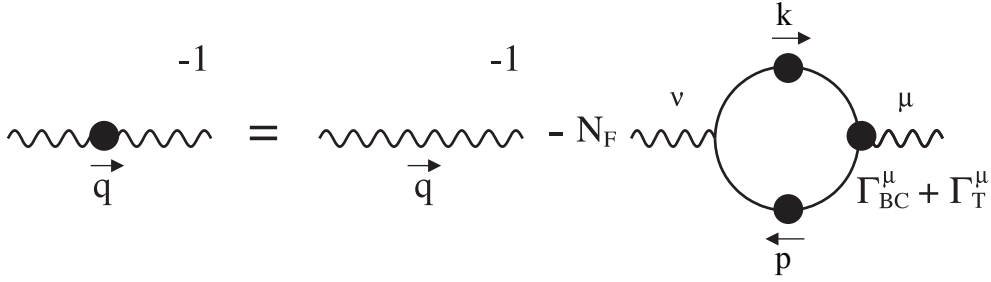} 
 \caption{Photon Schwinger-Dyson Equation.}
 \label{default}
 \end{center}
 \end{figure}

The renormalized photon SDE from Eq.~(\ref{eq:mainrensdph}) is
\begin{equation}
    \Delta^{-1}_{\mu\nu}(q) = Z_3(\mu)\,\left(\Delta^0_{\mu\nu}\right)^{-1}(q) + \,Z_1(\mu)\,{\overline{\Pi}}_{\mu\nu}(q)\,,
\label{eq:mainsdph2}
\end{equation}
where ${\overline{\Pi}}_{\mu\nu}$ is the photon vacuum polarization or self-energy obtained by evaluating the
photon SDE diagram using the Feynman rules.

Similar to the Ward-Green-Takahashi identity for fermion propagator,  
the Ward-Takahashi identity for the photon propagator is:
\begin{equation}
    q^\mu \, \Delta^{-1}_{\mu\nu} = \frac{q^\nu q^2}{\xi}\,.
    \label{eq:phwardtaka}
\end{equation}
\noindent
Making use of this identity, Eq.~(\ref{eq:phwardtaka}), for the photon SDE, Eq.~(\ref{eq:mainsdph2}), 
leads us to the well known {\it{transversality condition of the photon self-energy}}~:
\begin{equation}
    q^\mu \overline{\Pi}_{\mu\nu} = 0\,.
    \label{eq:transversality}
\end{equation}
If we contract the photon self-energy with $q^\mu$~:
\be
q^\mu {\overline{\Pi}}_{\mu\nu}(q)&=&i e^2 N_F \> {\rm Tr} \int_M {d^4 k \over (2\pi)^4}\> \gamma_\nu \>S(k) \>(q \cdot \Gamma(p,k)) \>  S(p)\,,
\ee
and use the WGTI 
\be
q^\mu {\overline{\Pi}}_{\mu\nu}(q)&=&i e^2 N_F \> {\rm Tr} \int_M {d^4 k \over (2\pi)^4}\> \gamma_\nu \>(S(k) -  S(p=k-q))\,.
\label{eq:qpimu}
\ee
One expects that this integral is trivially zero, since the integration variable in the second term can be shifted so that it cancels out the first term.  Although this is the case  if one employs a gauge-covariant regularization scheme such as Dimensional Regularization, it is not true for the cut-off regularisation  since this integral is linearly divergent and one is not allowed to perform any shift in the integration variable.
In a UV cut-off regularization scheme the bare quantities are not gauge invariant because the divergent integrals depend on the position of the 4-dimensional hypersphere defined by the cut-off, i.e., the divergent integrals are not invariant under momentum shifts. However, the great benefit of the regularization-independent approach is that by only calculating finite convergent quantities gauge invariance is restored as $\Lambda \longrightarrow \infty$. This occurs because the differences of the divergent bare integrands in Eq.(\ref{eq:qpimu}) give rise to convergent integrands which respect invariance under momentum shifts as $\Lambda \longrightarrow \infty$.

 Therefore if the  gauge invariance is respected by regularization method this term is transverse and finite.  
The transversality condition suggests that the photon self energy tensor must admit the following tensor decomposition~:
\begin{equation}
    \overline{\Pi}_{\mu\nu} = -q^2 \left(g_{\mu\nu}-\frac{q_{\mu} q_\nu}{q^2}\right) \, \overline{\Pi}(q^2)\,,
\label{eq:pimunu}
\end{equation}
where $\overline{\Pi}$ is called the scalar self-energy. Using the following transverse projector
\be
	P_{\mu\nu}=\frac{-1}{(d-1) q^2} \, \left(g_{\mu\nu}-d \,\frac{q_{\mu} q_\nu}{q^2}\right)\,,
\ee
with $d$ is the dimensionality of space-time, the inverse relation can be found as
\be
{\overline{\Pi}} = P^{\mu\nu} \, {\overline{\Pi}}_{\mu\nu}\,.
\label{eq:pibar}
\ee

\noindent
Inserting Eqs.(\ref{eq:invbareph}, \ref{eq:InvPhotonPropagator}) in Eq.(\ref{eq:mainsdph2}), imposing the transversality condition, Eq.~(\ref{eq:pimunu}) with Eq.(\ref{eq:pibar}),  and cancelling the longitudinal components and common factors, yields
the following equation for the inverse photon wave-function renormalization function, $G$~:

\be
\frac{1}{G(q^2,\mu^2)} &=& Z_3(\mu) + Z_1(\mu)\, {\overline{\Pi}(q^2)}\,.
\label{eq:G1}
\ee
From Eq.(\ref{eq:mainsdph}), the photon self energy tensor can be written explicitly as
\be
{\overline{\Pi}}_{\mu\nu}(q)&=&i e^2 N_F \> {\rm Tr} \int_M {d^4 k \over (2\pi)^4}\> \Gamma_\mu (p,k)\>S(k) \>\gamma_\nu \>  S(p)\,,\>\nonumber \\
    &=& i e^2 \, N_F \int_M  {\text{d}^4 k \over (2\pi)^4} \, \text{Tr}
    \left[
            \Gamma_\mu(p,k) \left(\kslash + M(k^2)\right) \gamma_\nu \left(\pslash + M(p^2)\right)
        \right] \nn\\ 
  && \hspace{3cm} \times\, \frac{1}{A(k^2)\left[k^2 - M^2(k^2)\right]} \, \frac{1}{A(p^2)\left[p^2 - M^2(p^2)\right]} \,. 
  \label{eq:pimunu2}
\ee

\vspace*{3mm}

\noindent
Using an analogous procedure to the fermion propagator in Eq.~(\ref{eq:submass}), we can form the
appropriate subtractions of the renormalized photon SDEs, Eq.(\ref{eq:G1}) to eliminate the divergent renormalization
constants $Z_1$ and $Z_3$ by recalling that $G(\mu^2;\mu^2) \>=\> 1$ yield~:

\be
\fbox{
\vspace*{0.5cm}
$ \displaystyle
G^{-1}(\mu^2;q^2)\> = \displaystyle \>1 \> + \> \left[ G^{-1}(\mu^2;q^2)\bar \Sigma_d(\mu^2) \> + \> \bar \Pi(q^2)\right] \> -
\>\left[ \bar \Sigma_d(\mu^2)\> + \> \bar \Pi(\mu^2) \right] \;\;\;.$}
\label{eq:subphoton}
\ee

\vspace{5mm}

\noindent
{\bf{\underline{G equation - The Photon Wave-function Renormalisation:}}}

Making use of Eqs.(\ref{eq:pibar},~\ref{eq:pimunu}), the photon self-energy  in Euclidean space can be written as~:

\begin{equation}
   \overline{\Pi}(q^2)
    = \frac{\alpha \, N_F}{3\pi^3} \int_E \text{d}^4 k \frac{1}{q^2} \,
        \frac{1}{A_p\,(p^2 + M_p^2)} \, \frac{1}{A_k\,(k^2 + M_k^2)}
        \, \Big\{ \mathcal{I}^{L}_\pi +  \mathcal{I}^T_\pi \Big\}\,,
	\label{eqn:photon_self_energy}
\end{equation}
where $N_F$ is the number of fermion flavors, $\mathcal{I}^{L}_{\pi} $ and $\mathcal{I}^{T}_{\pi} $ are the
integrands related to the longitudinal and transverse part of the fermion-photon vertex of the photon self energy, ${\overline{\Pi}}(q^2)$, respectively can be written~:
\be
\mathcal{I}^{L}_\pi
&=& \frac{1}{2} \left(A_k + A_p\right) \,
        \left[ 2 k \cdot p - \frac{8}{q^2}\left(\Delta^2+q^2 k \cdot p\right) \right] \nonumber\\[3mm]
&+& \frac{1}{2} \frac{\left(A_k - A_p\right)}{(k^2-p^2)} \,
        \bigg[ \left(-\left(k^2+p^2\right) + 2 M_k M_p \right)
                \left\{\frac{8}{q^2} \left(k\cdot q\right)^2 - 3 k \cdot q - 2 k^2 \right\} \nonumber \\[3mm]
&&              \hspace{2.8cm} -3 \left(k^2-p^2\right) \left(M_k M_p - k^2\right) \bigg] \nonumber \\[3mm]
&+&\frac{ \left(M_k\,A_k - M_p\,A_p\right)} {k^2-p^2} \,
        \bigg[ -\left(M_k + M_p \right)
                \left\{\frac{8}{q^2} \left(k\cdot q\right)^2 - 3 k \cdot q - 2 k^2 \right\} \nonumber\\
 &&\hspace{3.6cm}               +\,3 \left(k^2-p^2\right) M_k \bigg] \,,
\label{eqn:pi_L}               \\[5mm]
\mathcal{I}^{T}_\pi
&=& \tau_1^E(p^2,k^2,q^2) \,
    \left[ M_k \left\{\Delta^2\right\} + M_p \left\{\Delta^2\right\} \right] \nonumber \\
&+& \tau_2^E(p^2,k^2,q^2) \,
    \left[ \left(k^2+p^2\right) \left\{-\Delta^2\right\} + M_k M_p \left\{2\Delta^2\right\} \right] \nonumber \\
&+& \tau_3^E(p^2,k^2,q^2) \,
    \left[ 3 q^2 k \cdot p + 2\Delta^2 + M_k M_p \left\{3q^2\right\} \right] \nonumber \\
&+& \tau_4^E(p^2,k^2,q^2) \,
    \left[ M_k \left\{3 q^2 k \cdot p + 2\Delta^2 + 3 p^2 q^2 \right\}
      -M_p \left\{3 q^2 k \cdot p + 2\Delta^2 + 3 k^2 q^2 \right\} \right] \nonumber \\
&+& \tau_5^E(p^2,k^2,q^2) \,
    \left[ M_k \left\{ 3 p \cdot q \right\} - M_p \left\{ 3 k \cdot q \right\} \right] \nonumber \\
&+& \tau_6^E(p^2,k^2,q^2) \,
    \left[ 3k \cdot p \left(p^2 - k^2\right) + M_k M_p \left\{ 3\left(p^2-k^2\right)\right\} \right] \nonumber \\
&+& \tau_7^E(p^2,k^2,q^2) \,
    \bigg[ M_k \left\{ \frac{3}{2} \left(p^2-k^2\right) \left(p^2+k\cdot p\right) - \Delta^2 \right\} \nonumber\\
 &&  \hspace{2.5cm}   -M_p \left\{ \frac{3}{2} \left(p^2-k^2\right) \left(k^2+k\cdot p\right) + \Delta^2 \right\} \bigg] \nonumber \\
&+& \tau_8^E(p^2,k^2,q^2) \,
    \left[ -2\Delta^2 \right] \,.
    \label{eqn:pi_T}
\ee

To solve the fermion wave-function renormalisation, $F$, Mass function, $M$ in Eq.(\ref{eq:submass}) and the photon wave-function renormalisation, $G$, in Eq.(\ref{eq:subphoton}), simultaneously and analytically for a given vertex is not possible since 
these are nonlinear integral equations, unless one makes major approximations however it is possible to solve them numerically using numerical iteration methods.  

Unquenching the theory adds many challenges to this procedure and presents itself complications, nevertheless it is possible achieve this using advance numerical calculation techniques.  Below we will present our numerical results and discuss them in some detail.
 
\section{\label{sec:Numerical solutions} Numerical Solutions}

Now we turn to numerical solutions of the coupled equations for the unquenched fermion and photon propagators, Eqs.~(\ref{eq:subwavefunc}) and (\ref{eq:subphoton}), with self-energies given by Eqs.~(\ref{eqn:dirac_self_energy}),~(\ref{eqn:scalar_self_energy}) and (\ref{eqn:photon_self_energy}) respectively. The propagator functions $F(p^2) = 1/A(p^2)$, $M(p^2)$ and $G(p^2)$ take their values on a logarithmically-spaced grid of momentum-squared points covering 10-20 orders of magnitude, interpolated via cubic splines. The equations are iterated until they satisfy some convergence criteria; which we have chosen to be that the propagator functions vary from their previous incarnations at each point by less than 1 part in $10^6$.

Clearly this involves the introduction of both an infra-red cutoff $\lambda^2$ and an ultra-violet cutoff $\Lambda^2$ in the integrations of the self-energy equations:
\begin{equation}
        \int_0^{\infty} \text{d} k^2 \to \int_{\lambda^2}^{\Lambda^2} \text{d} k^2\,.
\end{equation}
A more complete analysis, would estimate the contribution from the IR and UV tails; but in the current context solutions meeting the convergence criteria are obtained by choosing $\lambda^2 \ll \Lambda^2$. Valid numerical solutions need to be stable against choice of momentum point density, and IR and UV cutoff (which is harder to achieve for massless solutions, where both $1/A$ and $1/G$ are infrared divergent).

%
%

\subsection{\label{sec:numerical solutions renormalized in from cutoff} Numerical solutions renormalized at $\mu^2 < \Lambda^2$}

In this section, we compare massless and massive solutions for various vertices (Bare, Ball-Chiu, Curtis-Pennington, Modified CP and Kizilersu-Pennington) which were identified above, with parameters UV cutoff $\Lambda^2 = 10^{12}$, IR cutoff $\lambda^2 = 10^{-2}$ and renormalization point $\mu^2 = 10^8$.

\begin{figure}[htbp!]
\begin{center}
 \includegraphics[width=12cm,angle=-90]{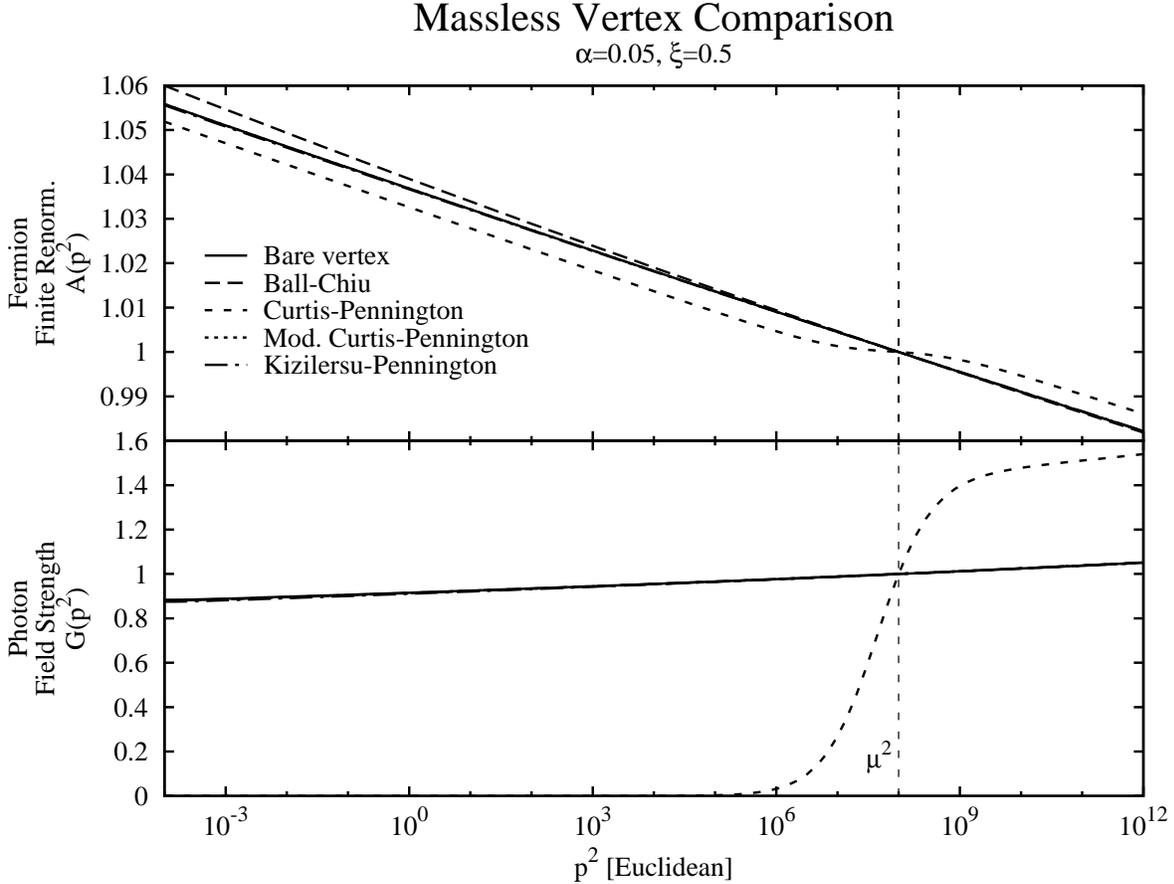}
 \caption{Massless Fermion and Photon wave-function renormalizations versus $p^2$ using Bare, BC, CP, mod. CP and KP vertices for $\alpha=0.05$ and $\xi=0.5$. The lines except CP and BC in above plot and CP in below plot overlaps. }
\label{fig:bare-bc-cp-kp-cmp}
\end{center}
\end{figure}

In Fig.~\ref{fig:bare-bc-cp-kp-cmp} for  gauge parameter $\xi=0.5$, we see that in order for the Curtis-Pennington vertex
to satisfy the photon SDE renormalization condition, the photon field strength is driven to zero in the IR.
Similar pathological behaviour for this vertex is exhibited in the Landau gauge; solutions for higher $\alpha$
do not even converge. This confirms that the ({\bf{unmodified}}) CP vertex has a dynamical problem in the unquenched photon
SDE which is also mentioned in Ref.~\cite{Bloch:1995dd} and so we eliminate it from further consideration in favour of the {\it{Modified Curtis-Pennington vertex}} which is described in Eq.(\ref{eq:modCP}).
%
%
%
%
%
\subsubsection{\label{sec:masslessvsmassive}Massless versus Massive Solutions}

\begin{figure*}[htbp]
\begin{center}
\includegraphics[width=12.5cm,angle=-90]{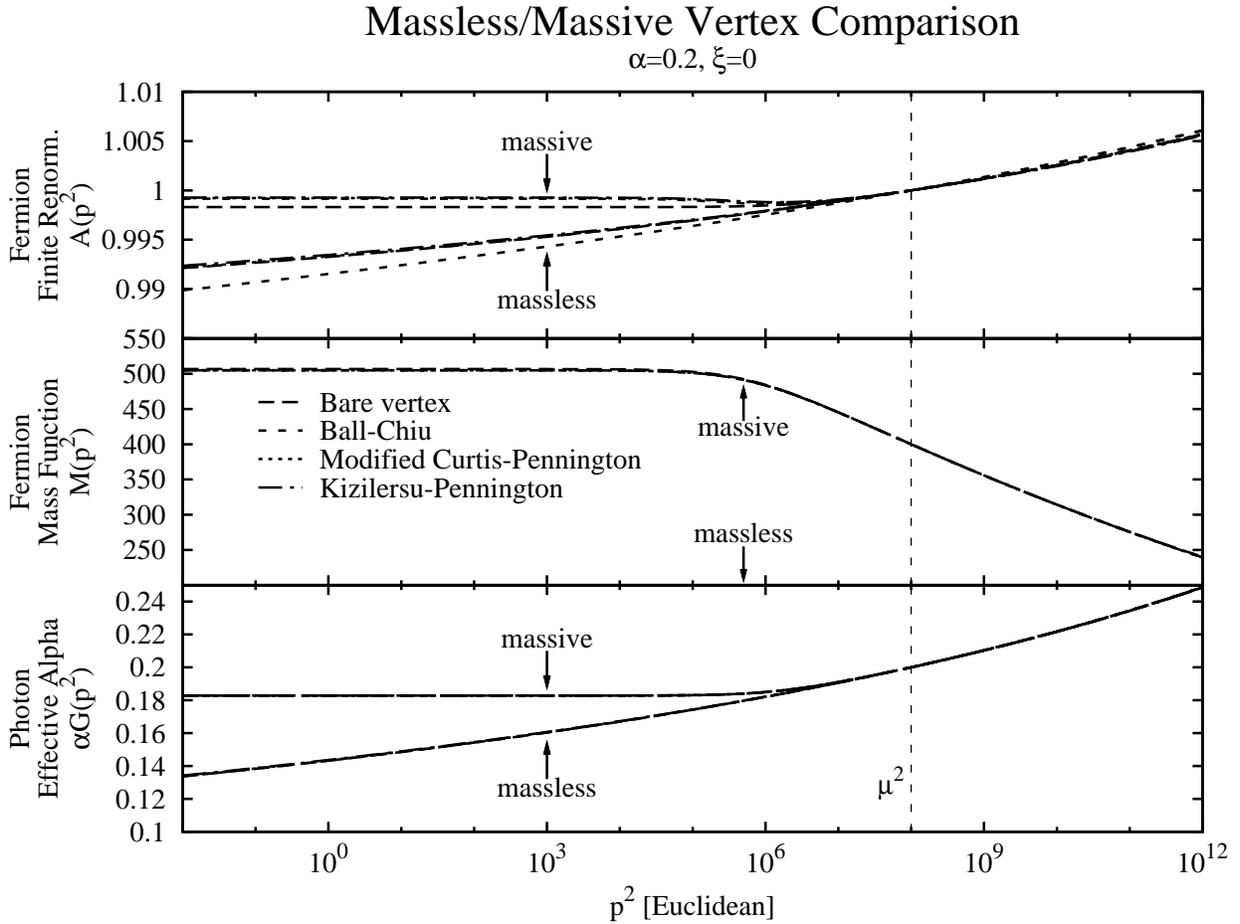}
\end{center}
\caption{Vertex comparison of massless and massive $\alpha=0.2$ solutions in Landau gauge. The massive solutions have $m_{\mu} = 400$.}
\label{fig:bare-bc-cp-kp-massless-massive-cmp0}
\end{figure*}

We compare the remaining vertices in the massless, ({\it{i.e. M identically zero, $M = 0$}}), and massive, $M \neq  0$, unquenched propagators in Landau gauge for $\alpha=0.2$ in Fig.~\ref{fig:bare-bc-cp-kp-massless-massive-cmp0}.
These solutions converge and are similar in form to the quenched solutions of $A$ and $M$ exhibited previously \cite{ Hawes:1991,Hawes:1996mw,Hawes:1996ig,Hawes:1994ce,Kizilersu:2001}. In particular, the massless $A$ and $G$ functions tend to zero in the IR, while their massive counterparts tail off in the IR to a non-zero constant. All vertices give similar results for the photon propagator, and for the mass function in the massive case. Only the fermion finite renormalization function $A(p^2)$ results differ: in the massless case, the BC vertex solution is appreciably different from the other vertex solutions; this gives credence to the view that the BC vertex needs to be supplemented by a transverse part to restore (and improve) the characteristic solution. On the other hand, in the massive case, the bare vertex solution differs from the others. Massless and massive solutions overlap in the asymptotic region.

\begin{figure*}[htbp]
\begin{center}
\includegraphics[width=12.5cm,angle=-90]{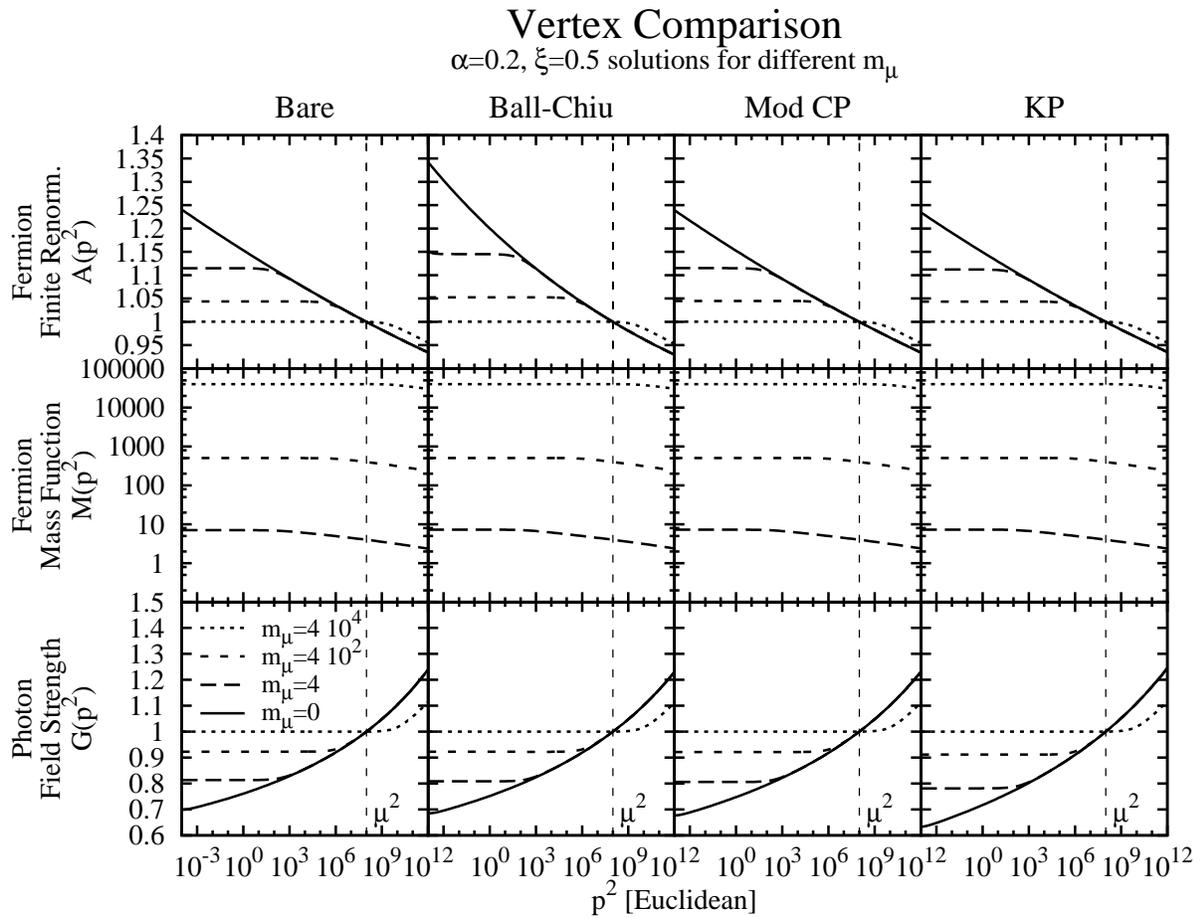}
\end{center}
\caption{Vertex comparison of the massive $\alpha=0.2, {\xi}=0.5$ solutions for different $m_\mu$ values.}
\label{fig:cmp4mass}
\end{figure*}

Figure~\ref{fig:cmp4mass} presents the same massless/massive comparison of the solutions in a different format for $\alpha=0.2$ and $\xi=0.5$. Again, we see that massive solutions of $A$ and $G$ for small $m_\mu$ share a common asymptotic tail with their massless counterparts.
We note that increasing $m_{\mu}$ results in the flat IR tail increasing towards higher momentum for all the propagator functions, but having the same UV tail, until $m_{\mu}^{2}$ exceeds $\mu^{2}$, whereupon the asymptotic tail itself is shifted towards higher momenta, while the flat IR tail is shared.

For the remainder of this section, we concentrate on massive solutions.
%
%
%
%
\subsubsection{\label{sec:KP vertex comparison}KP Vertex Comparison}

\begin{figure}[htbp!]
\begin{center}
 \includegraphics[width=12.5cm,angle=-90]{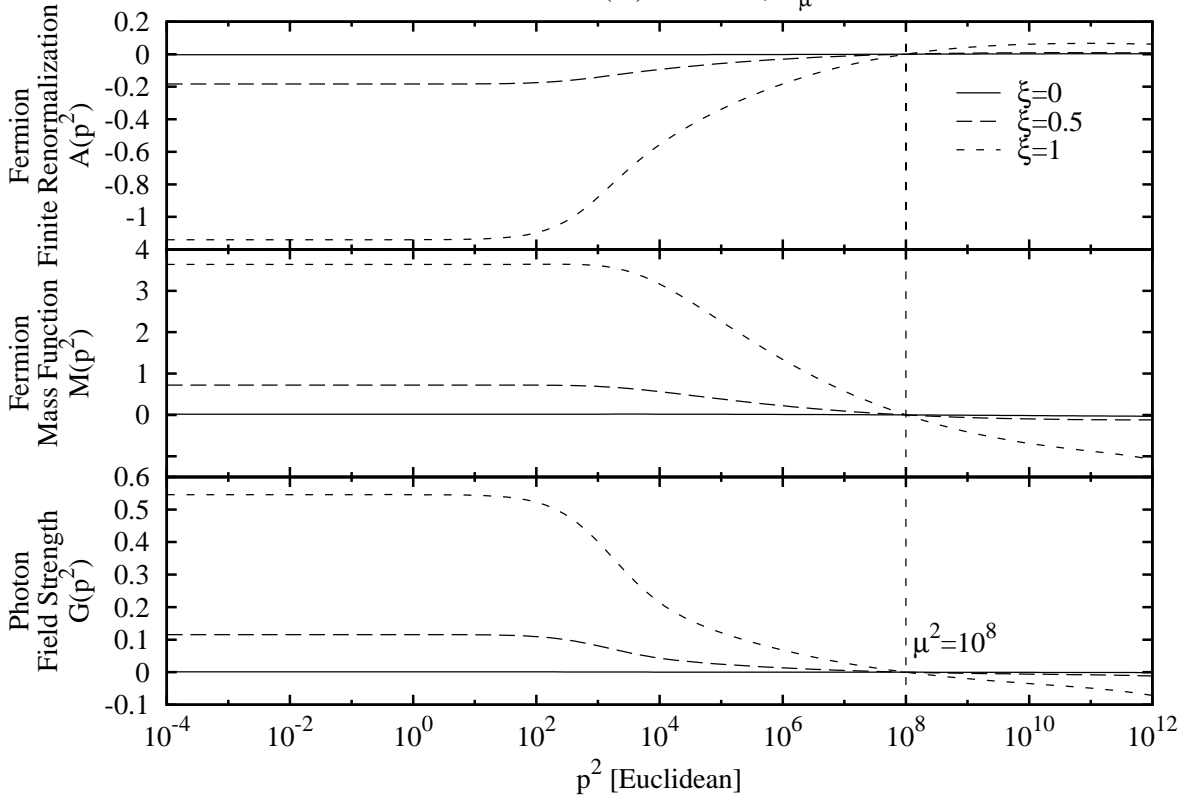}
 \caption{Percentage relative differences between the KP-Type2 and KP-Type3 in fermion, photon wave-function renormalizations and mass function versus $p^2$ using  vertices for $\alpha=0.6$ and $\xi=0, 0.5, 1.0$.}
\label{fig:kp-type-3-5.cmp}
\end{center}
\end{figure}

One of the purpose of this paper is to understand the effect of the KP vertex in the unquenched fermion and photon propagators. This vertex 
is formulated to satisfy the requirements of Multiplicative Renormalisability (MR) of both  the fermion and photon propagators and these constraints can be matched by few specific KP vertex constructions, Eqs.(\ref{eq:finaltaus3},~\ref{eq:finaltaus2}). These Type 2 and Type 3  KP vertices have the same unique photon limit 
$\protect{k^2 \simeq p^2 \gg q^2}$ and both satisfy all other necessary constraints but they differ from each other only beyond the leading logarithmic order
as explained in Ref.\cite{Kizilersu:2009kg}. Figure \ref{fig:kp-type-3-5.cmp} displays relative percentage difference in the solutions of $A$, $M$ and $G$ with the parameters 
 $\alpha=0.6$ and $\xi=0, 0.5,1.0$ for two types of KP vertices, Eqs.(\ref{eq:finaltaus3},\ref{eq:finaltaus2}).  They yield almost identical results for the Landau gauge but in the Feynman gauge the difference between them is much more evident, and greater in the IR region than in the  UV region for the renormalization point, $(\mu^2=10^8)$, obviously this conclusion may change according to the chosen renormalization point.

%
%
%

\subsubsection{\label{sec:quenchedvsunquenched}Quenched versus Unquenched Solutions}

\begin{figure*}[htbp]
\begin{center}
\includegraphics[width=12.5cm,angle=-90]{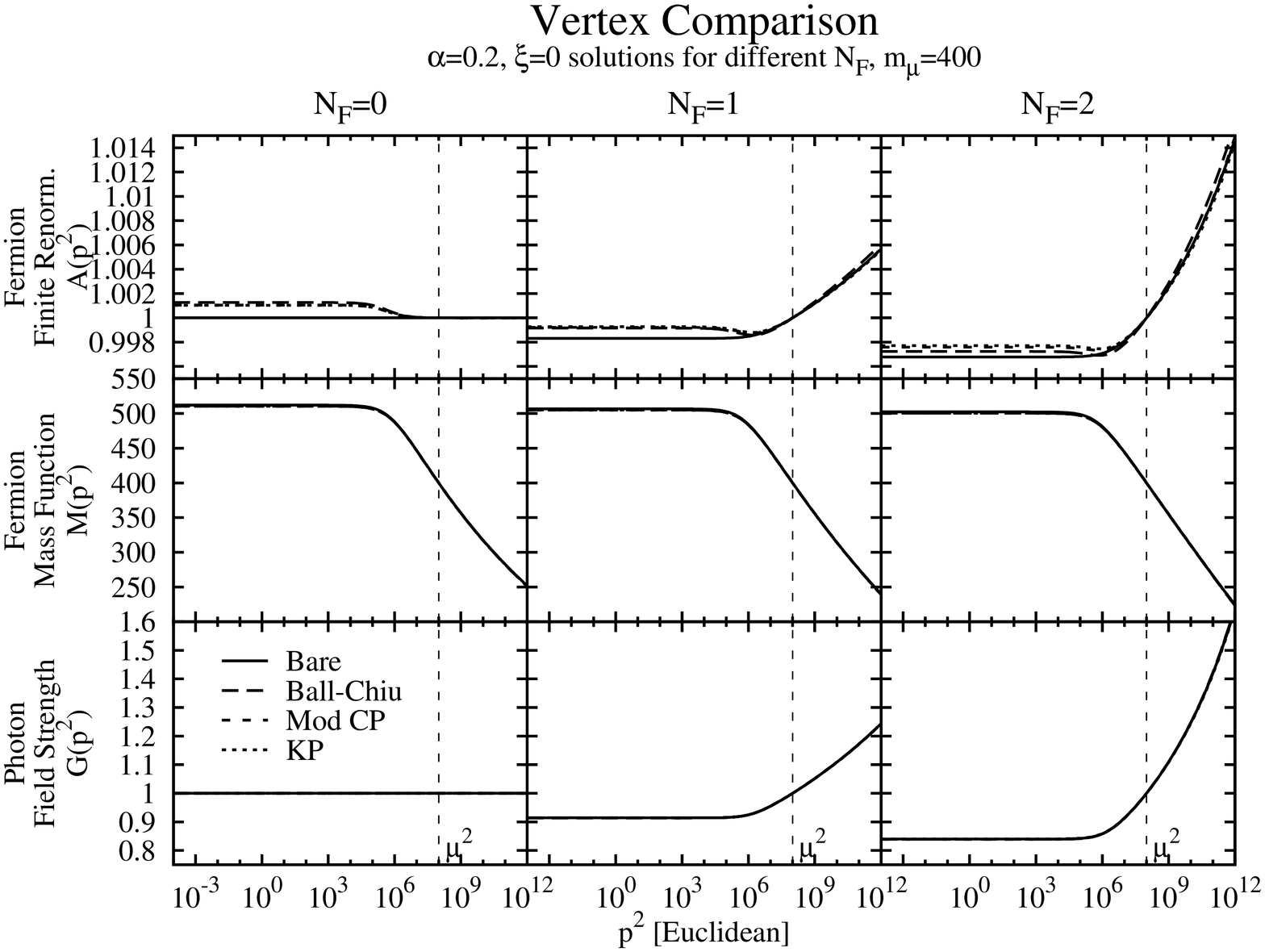}
\end{center}
\caption{Massive Bare, BC, modified CP and KP Vertex comparisons:  $\alpha=0.2, \xi=0 $ solutions for different  $N_F$ , $m_{\mu}=4$.}
\label{fig:cmp4massnflav0}
\end{figure*}

\begin{figure*}[htbp]
\begin{center}
\includegraphics[width=12.5cm,angle=-90]{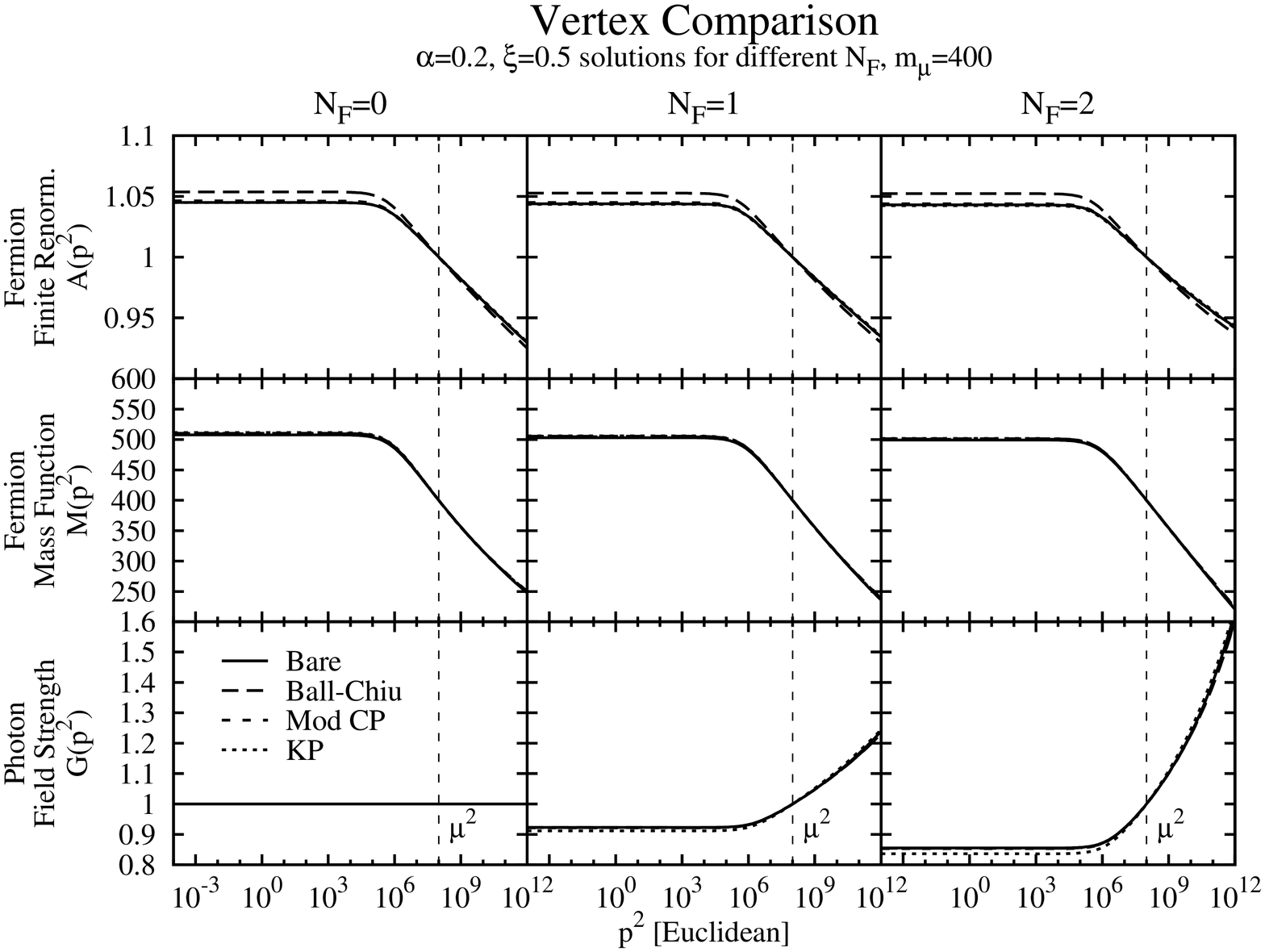}
\end{center}
\caption{Massive Bare, BC,modified CP and KP  Vertex comparisons:  $\alpha=0.2, \xi=0.5 $ solutions for different  $N_F$ , $m_{\mu}=4$.}
\label{fig:cmp4massnflav}
\end{figure*}

Figures~\ref{fig:cmp4massnflav0} and \ref{fig:cmp4massnflav} explore the effect of varying the number of flavours between $N_F=0,1,2$ for $\alpha = 0.2$ and $\xi=0$ and $0.5$ respectively. $N_F=0$ corresponds to the quenched case and $N_F=1$ represents the default case used in the other graphs. As expected, the primary effect of the variation is on the photon propagator $G$; its (indirect) effect on the fermion propagator functions is small, except for $A$ function in Landau gauge, which is close to unity and thus more susceptible to variation. 

Recall from Ref.\cite{ Kizilersu:2001} that the asymptotic form of the quenched fermion propagator functions have a (real or complex) power-law functional form
as a consequence of the scale invariance of the quenched theory (which follows from the renormalization group equations with constant coupling). By contrast, the unquenched solutions are not asymptotically power-law behaved, as a consequence of fermion loops in the photon SDE destroying the scale invariance of the unquenched theory.

\begin{figure*}[htbp]
\begin{center}
\includegraphics[width=12.5cm,angle=-90]{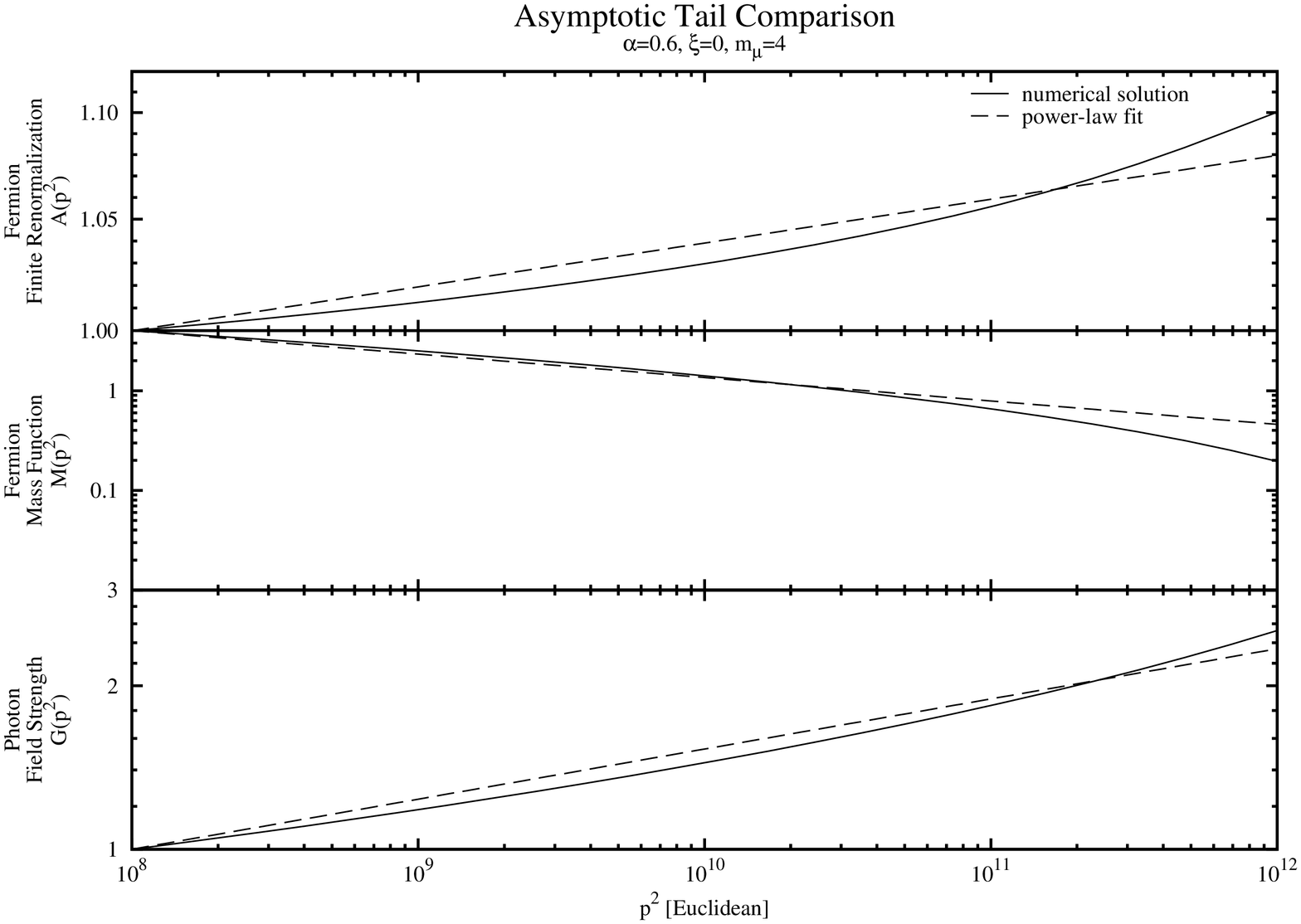}
\end{center}
\caption{Asymptotic tail comparison of numerical solutions for $\alpha=0.6$, $\xi=0$, $m_{\mu}=4$ (solid line) and their power-law fit (dashed line). }
\label{fig:kp-pow-fit}
\end{figure*}

Figure~\ref{fig:kp-pow-fit} shows the asymptotic tail comparison between the numerical solutions and the power-law fit. One can see that 
the solutions do not admit a power law behaviour for the unquenched solutions suggesting that they are different from their quenched partners. 
%
%
%

\subsubsection{\label{sec:gaugedependence}Gauge Dependence of the Solutions}

\begin{figure*}[h]
\begin{center}
\includegraphics[width=12.5cm,angle=-90]{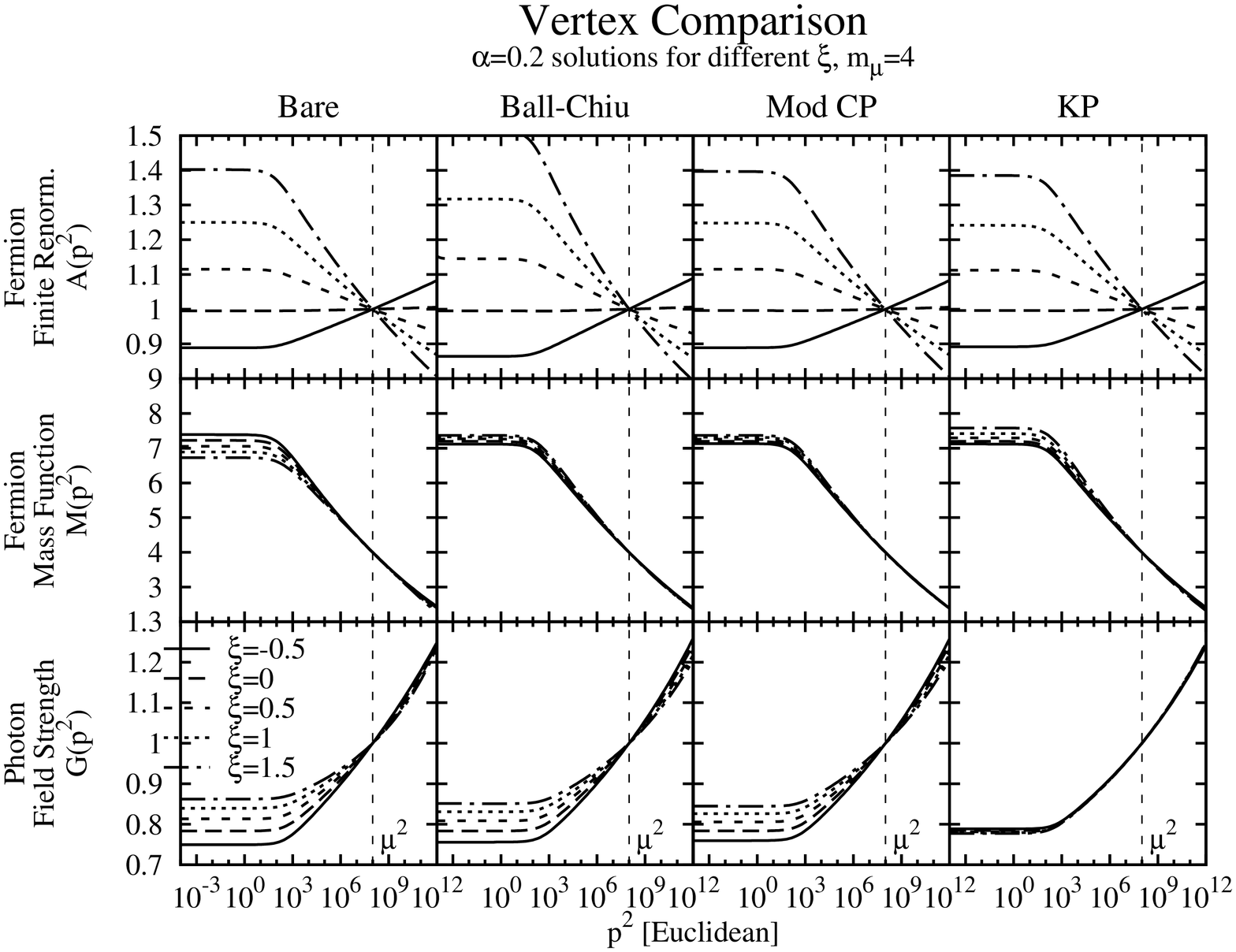}
\end{center}
\caption{Massive Bare, BC, modified CP and KP vertex comparisons:  $\alpha=0.2 $ solutions for different  $\xi$ , $m_{\mu}=4$.}
\label{fig:cmp4ximass}
\end{figure*}

In QED  {\it the photon propagator $G(p^2)$ should be gauge invariant, i.e. independent of $\xi$, while the fermion propagator should depend on gauge parameter in accordance with the Landau-Khalatnikov equations} \cite{LK}. The effect of varying the gauge parameter on the different vertices is explored in Fig.~\ref{fig:cmp4ximass}, where $\xi$ is varied between $-0.5$ and $1.5$ for $\alpha = 0.2$. The photon wave-function renormalization function should have no dependence on the gauge parameter. However note that {\it{$G(p^2)$ exhibits large variation with $\xi$ for all except the KP vertex}}, which approaches the desirable goal of gauge-independence.

%
%

\subsubsection{\label{sec:cutoffdependecd}Cut-off Dependence of the Solutions}
\begin{figure*}[h]
\begin{center}
\includegraphics[width=12.5cm,angle=-90]{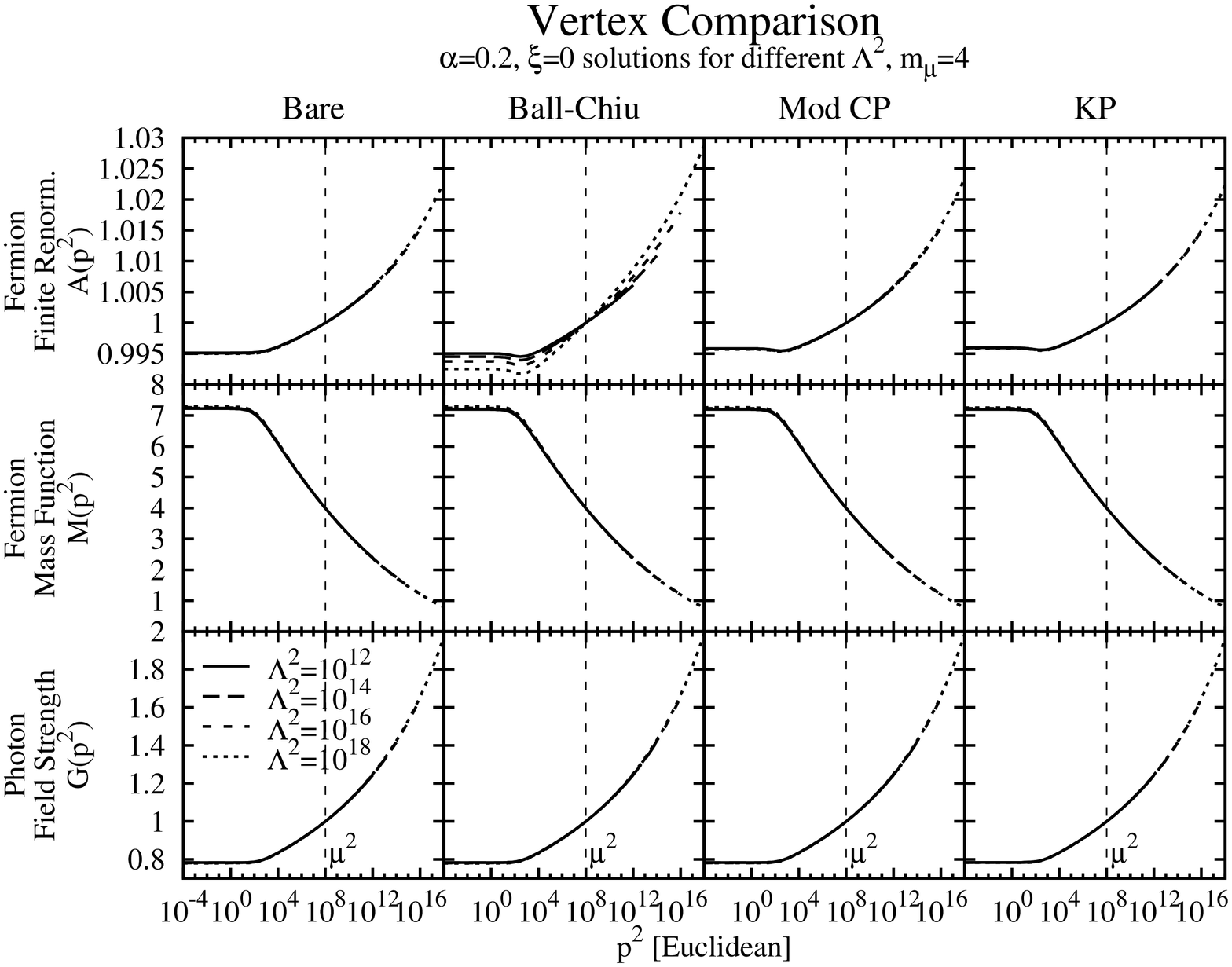}
\end{center}
\caption{Massive Bare, BC, modified CP  and KP Vertex comparisons:  $\alpha=0.2, {\xi}=0 $ solutions for different  $\Lambda^2$ , $m_{\mu}=4$.}
\label{fig:cmp4massxuv}
\end{figure*}
In Fig.~\ref{fig:cmp4massxuv}, we increase the cutoff $\Lambda^2$ from $10^{12}$ (the default case) to $10^{18}$ for $\alpha = 0.2$ and $\xi = 0$. For these renormalized solutions, we expect very little sensitivity to the choice of UV cutoff and, for all except the BC vertex, this is the case. Notably, the $A$ function of the BC vertex does vary appreciably with $\Lambda^2$ and we therefore exclude this vertex from further consideration.

%
%

\subsubsection{\label{sec:multiplicative renormalizability}Multiplicative Renormalizability of the Solutions}

\begin{figure*}[h]
\begin{center}
\includegraphics[width=12.5cm,angle=-90]{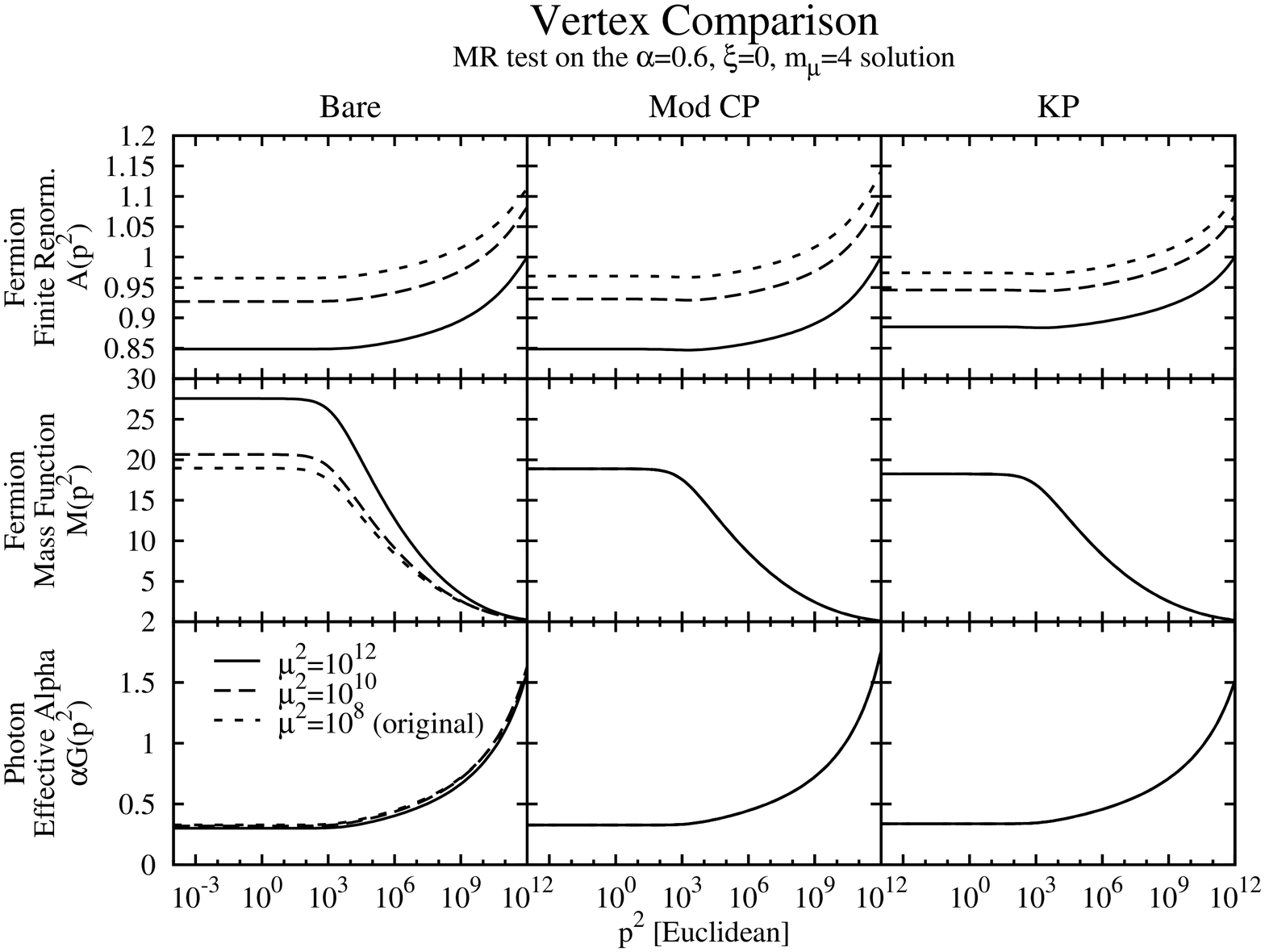}
\end{center}
\caption{Test of multiplicative renormalizability on $\alpha=0.6, {\xi}=0, m_{\mu}=4$ solutions.}
\label{fig:cmp3massmu2}
\end{figure*}
The propagator functions should depend on the renormalization point $\mu^2$ according to the renormalization group equations (RGE). We call this property the {\it MR test}. For QED, this is simplified by the observation that both {\it{ the mass function $M(p^2;\mu^2)$ and effective alpha $\alpha_{eff} = \alpha_\mu G(p^2;\mu^2)$ are renormalization-group invariants}} and should be unaffected by shifts in the renormalization point; $A$ should scale instead. Hence, given a solution set $A(p^2;\mu^2)$, $M(p^2)$ and $G(p^2;\mu^2)$ renormalized at $\mu^2$, new values for $\alpha$ and $m$ may be looked up from the effective alpha and mass functions respectively at the new renormalization point $\mu'^2$. The remaining boundary conditions are determined by the need for $A$ to be normalized at the new renormalization point, and for $\xi$ to transform oppositely to $\alpha$.
So the new (primed) solutions are related to the old (unprimed) solutions by
\be
        M'(p^2)         &=& M(p^2)    \,,          \\
        G'(p^2)         &=& \frac{\alpha}{\alpha'} G(p^2) =  G(p^2) / G(\mu'^2)\,, \\
        A'(p^2)         &=& A(p^2) / A(\mu'^2)\,,
\ee
where the (un)primed functions are functions of the (un)primed parameters respectively, for example,
\be
        A(p^2)  &=& A(p^2; \mu^2, \alpha_\mu, \xi_\mu, m_\mu)\,, \\
        A'(p^2) &=& A'(p^2; \mu'^2, \alpha_{\mu'}, \xi_{\mu'}, m_{\mu'})\,,
\ee
and the new parameters are related to the old by
\be
        m'              &=& M(\mu'^2)  \,, \\
        \alpha'         &=& \alpha_{eff}(\mu'^2) = \alpha G(\mu'^2) \,,  \\
        \xi'            &=& \xi / G(\mu'^2) \, .
\ee

The multiplicitive renormalizability of the remaining vertices is explored in Fig.~\ref{fig:cmp3massmu2}, which applies the $\mu^2$ test, as explained above, on the $\alpha=0.6$, $\xi=0$, $\Lambda^2=10^{12}$ solution with mass $\protect{m_{\mu}=4}$ at the renormalization point $\mu^2=10^8$. 
For the Bare vertex, the mass function and effective coupling $\alpha$ of the calculated solutions renormalized at $\mu^2=10^{10}$  and $\mu^2=10^{12}$ ( which coincides with the cut-off same, $\Lambda^2=10^{12}$) do not coincide with the original solution; hence this vertex fails the MR test, and is not multiplicatively renormalizable. By way of contrast,  KP vertex pass the MR test, reflecting their multiplicatively renormalizability by construction moreover the modified CP vertex which is CP vertex for the fermion SDE equation makes the mass function  multiplicatively renormalizable as it is designed however photon SDE which employs BC vertex is also unexpectedly makes photon propagator multiplicatively renormalizable as well.

%
%
%
%

\subsubsection{\label{sec:Couplingdependence} Coupling Strength Dependence of the Solutions}

\begin{figure*}[h]
\begin{center}
\includegraphics[width=12.5cm,angle=-90]{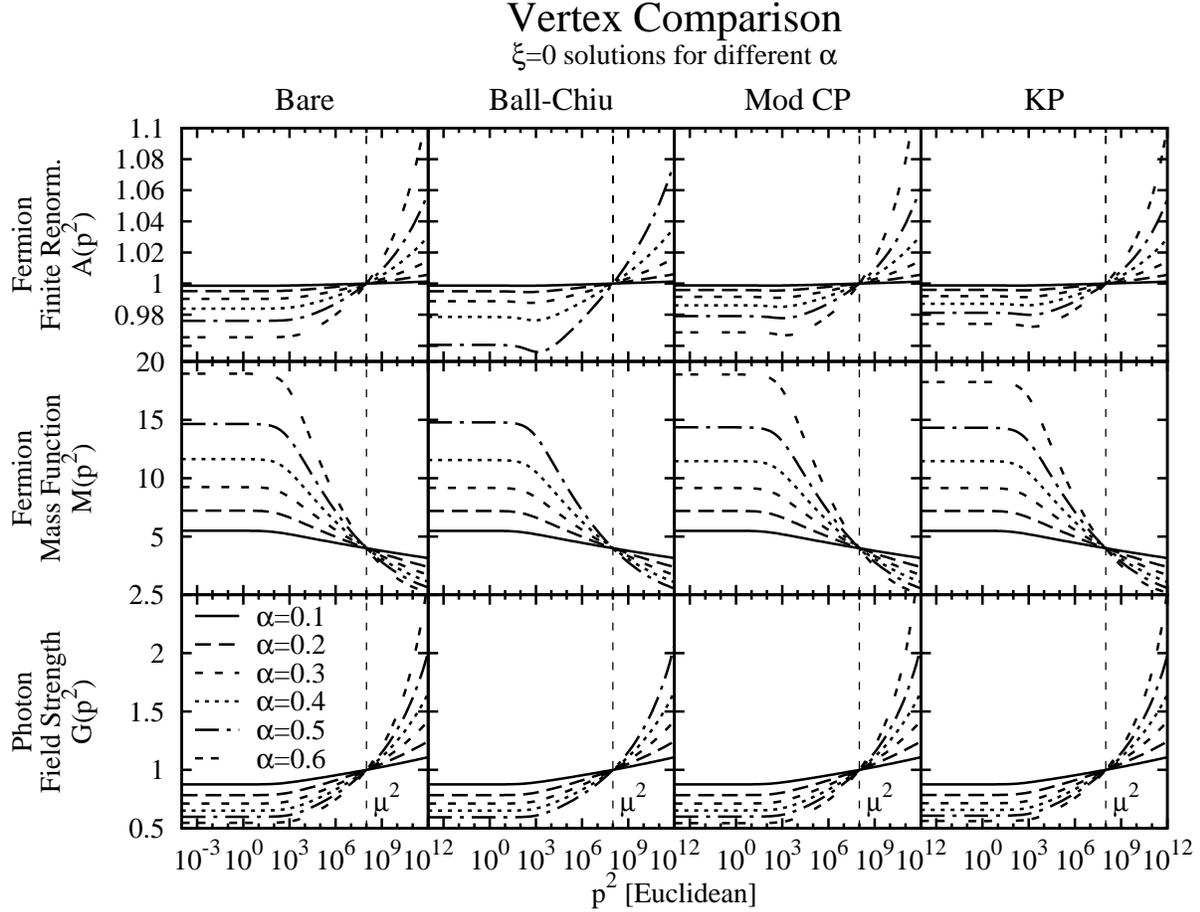}
\end{center}
\caption{Massive Bare, BC, modified CP and KP  vertex comparisons:  $\xi=0 $ solutions for different  $\alpha$, $m_{\mu}=4$.}
\label{fig:cmp4massalpha}
\end{figure*}
In Fig.~\ref{fig:cmp4massalpha}, $\alpha$ is varied between $0.1$ and $0.6$ in Landau gauge for the four vertices under consideration. As $\alpha$ increases the infrared values of $A$, $G$ decrease while $M$ increases and vice versa  in the UV region.  Note that the BC solution for $\alpha = 0.6$ does not converge, and is absent from the figure. The solutions for $\alpha > 0.6$ don't converge for any vertex choice (the limit seems to be just below $\alpha = 0.7$);  this appears to be a consequence of the existence of 
a zero of $1/G(p^2)$ -a {\it Landau pole}. This has fatal consequences for the integrand of the fermion propagator, which is proportional to $\alpha_{eff}(k^{2}) = \alpha_\mu G(k^2;\mu^2)$.

To establish an approximate limit on $\alpha$ imposed by the Landau pole, we consider
the first order leading log solution for $G$~:
\begin{equation}
        \frac{1}{G(p^2)} = 1 - \frac{\alpha_\mu \, N_F}{3 \pi} \ln \left(\frac{p^2}{\mu^2}\right) {\color{green}.}
        \label{eq:landau-pole-1}
\end{equation}
Since $G$ occurs within the Dirac and scalar self-energy integrands, Eqs.~(\ref{eqn:sigmad_L},\ref{eqn:sigmad_T}) and Eqns.~(\ref{eq:sigmas_L},\ref{eq:sigmas_T}) respectively, as a function of $q^{2}$, the maximum momentum evaluated in $G$ is at $p^2=4\Lambda^2$, neccessitating an extrapolation beyond the cutoff). So
\begin{equation}
         1/G(p^2) > 0 \quad\Rightarrow\quad \alpha_\mu \, N_F < \frac{3 \pi} {\ln(4\Lambda^2/\mu^2)} \,.
         \label{eq:landau-pole-2}
\end{equation}
Note that this dependends on the $\Lambda^{2}/\mu^{2}$ {\em ratio}.

\noindent
Choosing $\mu^{2}=10^{8}$ and $\Lambda^{2}=10^{12}$ as they were used in our prior analysis in this section, 
we find the limit from Eq.~(\ref{eq:landau-pole-2}) is 
\begin{equation}
	\alpha \, N_F < 0.89	\,.
	\label{eq:landau-pole-limit2} 
\end{equation}

\noindent
On the other hand, for solutions renormalized at the cutoff, $\mu^2=\Lambda^2=10^{12}$,  the limit imposed by  Eq.~(\ref{eq:landau-pole-2}) is 
\begin{equation}
	\alpha \, N_F < 6.8	\,.
\label{eq:landau-pole-limit1}
\end{equation}
The actual asymptotic behaviour and settling down to convergence invariably lowers this limit.
%
%
%

To summarize, the following defects of the vertices in our detailed study were noted: 
\begin{itemize}
\item the CP vertex is not dynamically viable,
\item the BC vertex is not invariant against the cutoff,
\item the bare vertex (and also, we expect, the BC vertex) is not multiplicatively renormalizable, 
\item all except the KP vertex have strongly gauge dependent photon propagators. 
\end{itemize}

Moreover, in contrast to the quenched case, we cannot advance the coupling strength $\alpha$ beyond fairly modest limits for any vertex choice. We summarise this behaviour in Table \ref{table:vertex comparison}. 

\begin{table}[htdp]
\caption{Vertex Comparison}
\begin{center}
\begin{tabular}{|l|c|c|c|c|}
\hline
 Vertices/Properties                                             & Bare Vertex      &    Ball-Chiu     & \, Mod. CP\,\,     & \,\,\,KP\,\,\, \ \\
\hline
 Invariance against $\Lambda^2$                          & $\surd$             &  $\times$       & $\surd$              & $\surd$  \\
\hline
 Multiplicative Renormalization Test                      & $\times$           &  $\times$       &  $\surd$             & $\surd$ \\
\hline
 Gauge Independence of Photon Propagator              &  $\times$          &   $\times$      & $\times$            & $\surd$   \\
\hline
\end{tabular}
\end{center}
\label{table:vertex comparison}
\end{table}

\subsection{\label{sec:numerical solutions renormalized at cutoff} Numerical solutions renormalized at $\mu^2 = \Lambda^2$.}

%
%

To overcome the limitation on $\alpha$ seemingly imposed by the existence of the Landau pole which we discussed in the last section, we explore solutions renormalized at the cutoff, which imposes the less draconian limit on Eq.~(\ref{eq:landau-pole-2}) given by Eq.(\ref{eq:landau-pole-limit1}). We use a higher cutoff $\Lambda^{2}=10^{14}$ so that the high-momentum behaviour of $G$ is more visible, but this is largely irrelevant for these {\em renormalized} solutions - as emphasised before, what is important is the $\Lambda^{2}/\mu^{2}$ ratio. Although the renormalized $\alpha$ increases with momentum, we can shift the renormalization point back to study the effect of the Landau pole on the solutions studied previously.

In this section, we study massless and massive solutions for the Bare, Modified CP and KP vertices renormalized at the cutoff $\Lambda^{2}=10^{14}$ mindful of the limits imposed by Eq.(\ref{eq:landau-pole-2}).
\begin{figure*}[htp]
\begin{center}
\includegraphics[width=12.5cm,angle=-90]{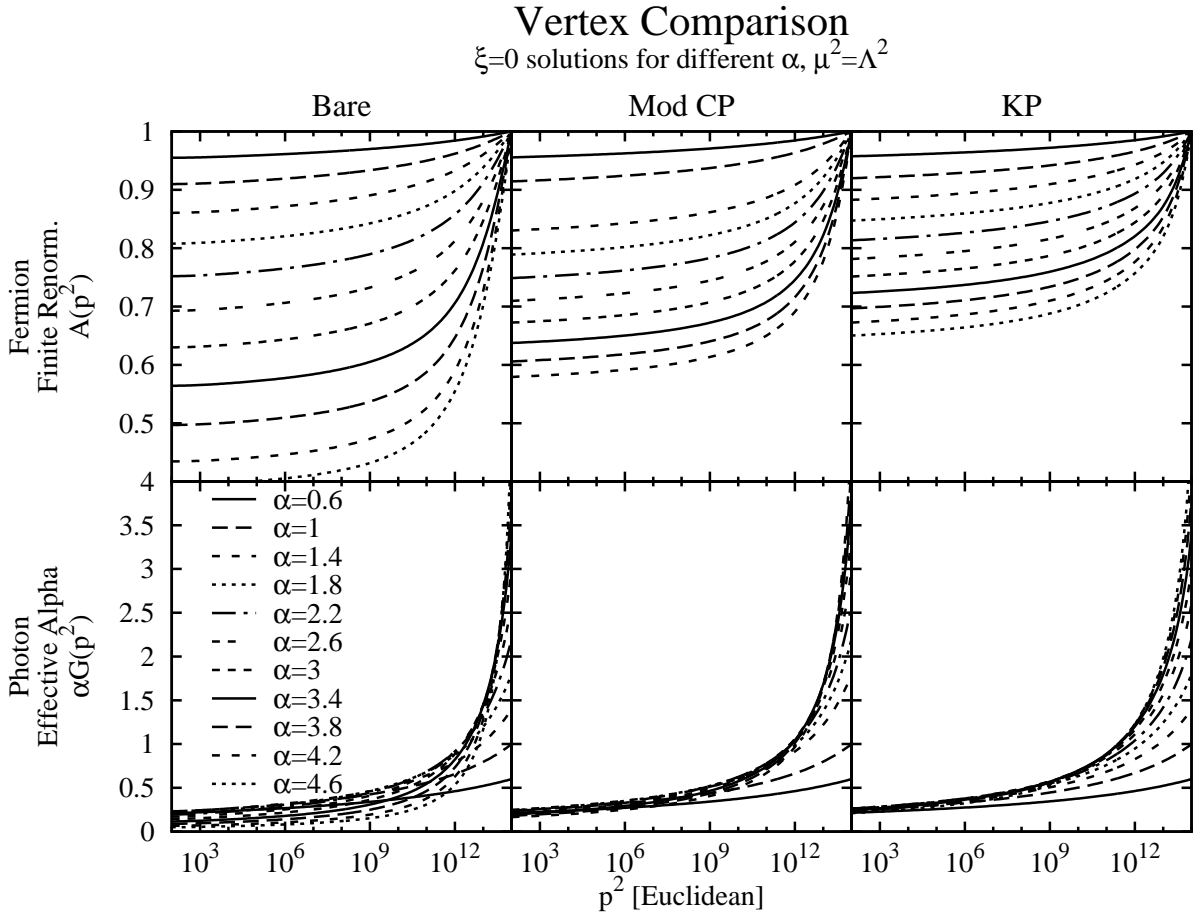}
\end{center}
\caption{Massless Bare, modified CP and KP Vertex comparisons:  $\xi=0 $ solutions for different  $\alpha$ ,  $\mu^2={\Lambda}^2$.}
\label{fig:cmp3cutalpha}
\end{figure*}

\begin{figure*}[htp]
\begin{center}
\includegraphics[width=12.5cm,angle=-90]{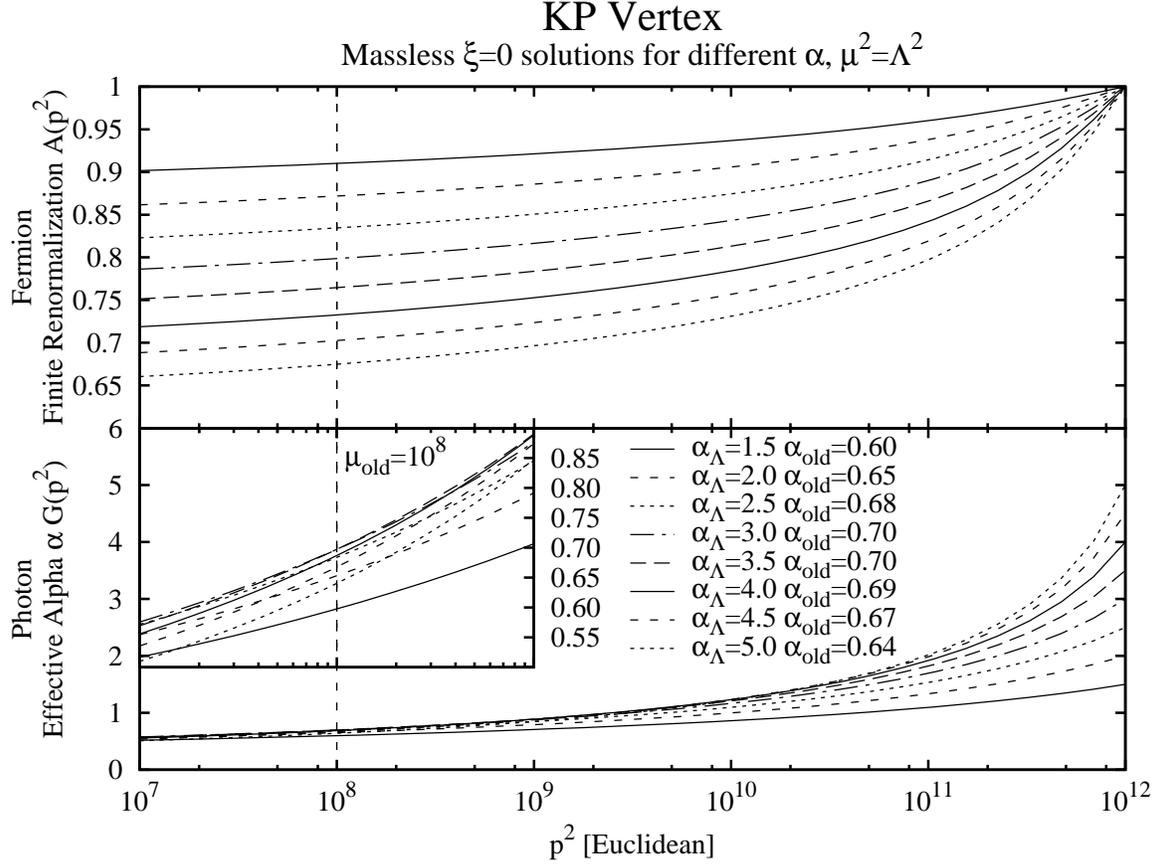}
\end{center}
\caption{Massless high $\alpha$ solutions for KP vertex.}
\label{fig:kp-sub-limit-alpha-ins}
\end{figure*}

Figure~\ref{fig:cmp3cutalpha} shows the effect of varying $\alpha$ between $0.6$ and $4.6$ for massless solutions renormalized at the cutoff. That the upper limit of $\alpha=4.6$ in the numerical studies is lower than the leading log approximations (which predicts $\alpha=6.8$) indicates the presence of higher order effects. In Fig.~\ref{fig:kp-sub-limit-alpha-ins}, we zoom in on the momentum range to see what happens to $\alpha$ around $p^2=10^{8}$, the ``old'' renormalisation point used in the previous section. The limit on $\alpha$, noted earlier is still very evident here, in fact, as  $\alpha_{\Lambda}$ is increased, 
$\alpha_{old}$ increases to $0.7$ but then decreases again! Hence, there exist the possibility of more than one solution satisfying the renormalization boundary conditions (for example, when specifying $\alpha_{\mu}$ between $0.6$ and $0.7$ for $\mu^{2}= 10^{8}$). Clearly, one of these degenerate solutions cannot satisfy the MR test: in practice, a `high' alpha solution reverts to the `low' alpha solution when used as a guess with the renormalization point set back from the cutoff.
%
%
%
%
\begin{figure*}[h]
\begin{center}
\includegraphics[width=12.5cm,angle=-90]{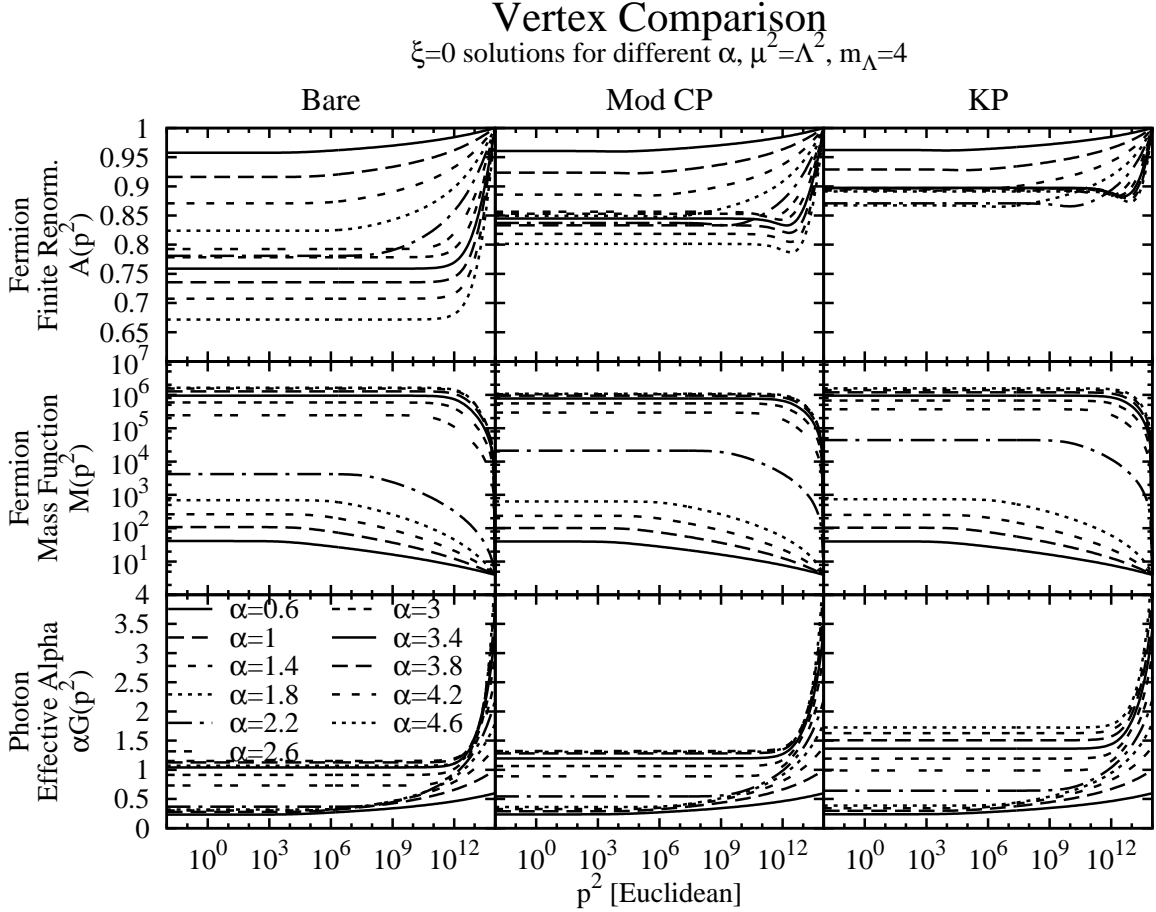}
\end{center}
\caption{A comparison of the Bare, Modified CP and KP vertices for $\xi = 0, \mu^2 = \Lambda^2$  for different $\alpha$'s over a range of $p^2$ values.}
\label{fig:cmp3cutalphamass}
\end{figure*}

By way of contrast, the massive solutions (with $m_\Lambda = 4$) renormalized at the cutoff presented in Fig.~\ref{fig:cmp3cutalphamass} and asymptotic to the corresponding massless solutions in $A$ and $G$) do not exhibit this limiting effect, but rather seem to separate into two distinct bands, which we label ``{\it{low alpha}}'' and ``{\it{high alpha}}''. The low alpha solutions correspond to those studied in the previous section. The high alpha solutions differ from the low alpha solutions in two ways~: 1)~$\alpha$ can now exceed the Landau pole limit at momenta less than the cutoff,  2)~the mass function is vastly amplified.

%
%

\begin{figure}
\begin{center}
\includegraphics[width=12.5cm,angle=-90]{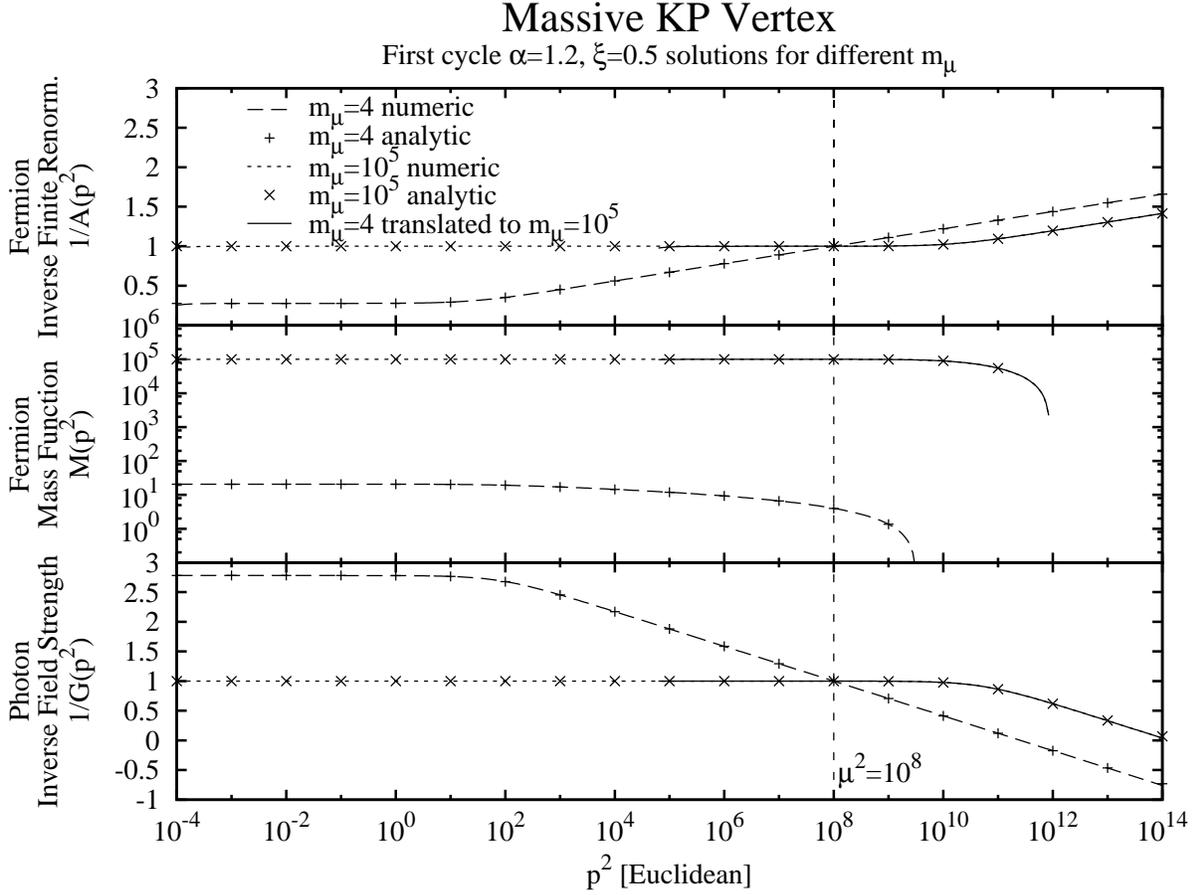}
\end{center}
\caption{Cycle 1 solutions for low and high mass from the numerical and analytic calculations.}
\label{fig:low-high-mass}
\end{figure}

To account for how the high alpha solutions seemingly evade the Landau pole limits, we study the first iteration cycle behaviour of the solutions with the KP vertex for $\alpha=1.2$ and $\xi=0.5$ renormalized at $\mu^2=10^{8}$ with cut-off $\Lambda^2 = 10^{14}$ for two choices of mass;  $m_{\mu}=m_{\rm low}=4$ and $m_\mu=m_{high}=10^5$ which are correspond to  ``{\it low mass}'' and ``{\it high mass}'' solutions respectively. The results are presented in Fig.~{\ref{fig:low-high-mass}. 

We do not expect the low mass solution to converge, and indeed in the first iteration cycle, the inverse photon propagator goes negative! On the other hand, the high mass solution remains viable in the first iteration cycle, by effectively translating the graph of low mass inverse photon propagator and the Landau Pole  to the right by a factor $(m_{high}/m_{low})^{2}$ as well as downwards as can be seen in Fig.~{\ref{fig:low-high-mass}. The other propagators are shifted similarly.

The equations resulting from the first iteration (which are independent of vertex) may be integrated analytically, the expressions are presented in Appendix \ref{sec:first cycle}. Therein, it is shown that the propagator functions $1/A(=F), M$ and $G$ at the low and high masses are related to each other according to two step process. Step 1 scales the solutions according to the relations below~: 
\begin{eqnarray}
F(p^2,\mu^2,m_{\rm{high}}^2) &=& F(p^2/s^2,\mu^2/s^2,m_{\rm{low}}^2)\,,\\
M(p^2,\mu^2,m_{\rm{high}}^2) &=& s\,\times\,M (p^2/s^2,\mu^2/s^2,m_{\rm{low}}^2)\,,\\
G(p^2,\mu^2,m_{\rm{high}}^2) &=& G (p^2/s^2,\mu^2/s^2,m_{\rm{low}}^2)\,,
\end{eqnarray}
where $s^2=m_{\rm{high}}^2/m_{\rm{low}}^2$.
\noindent
However during this scaling procedure the renormalization point of the propagator functions changes by an amount of $s^2$. Therefore the Step 2 process involves  obtaining the high mass solutions at the original renormalisation point from the scaled solutions. This procedure is explained in the Appendix~\ref{sec:first cycle}. 
 Figure~{\ref{fig:low-high-mass} shows the exceptional agreement between analytic and numeric evaluation of the first-iteration (cycle) of the solutions, as well as  the result of translating the low mass solution to the high mass solution.

{\it{From this we can conclude that the solutions with high $\alpha$ may exist if the mass is high enough; the mass function modifies the photon propagator so that it evades the Landau pole. }}
\begin{figure}
\begin{center}
\includegraphics[width=12.5cm,angle=-90]{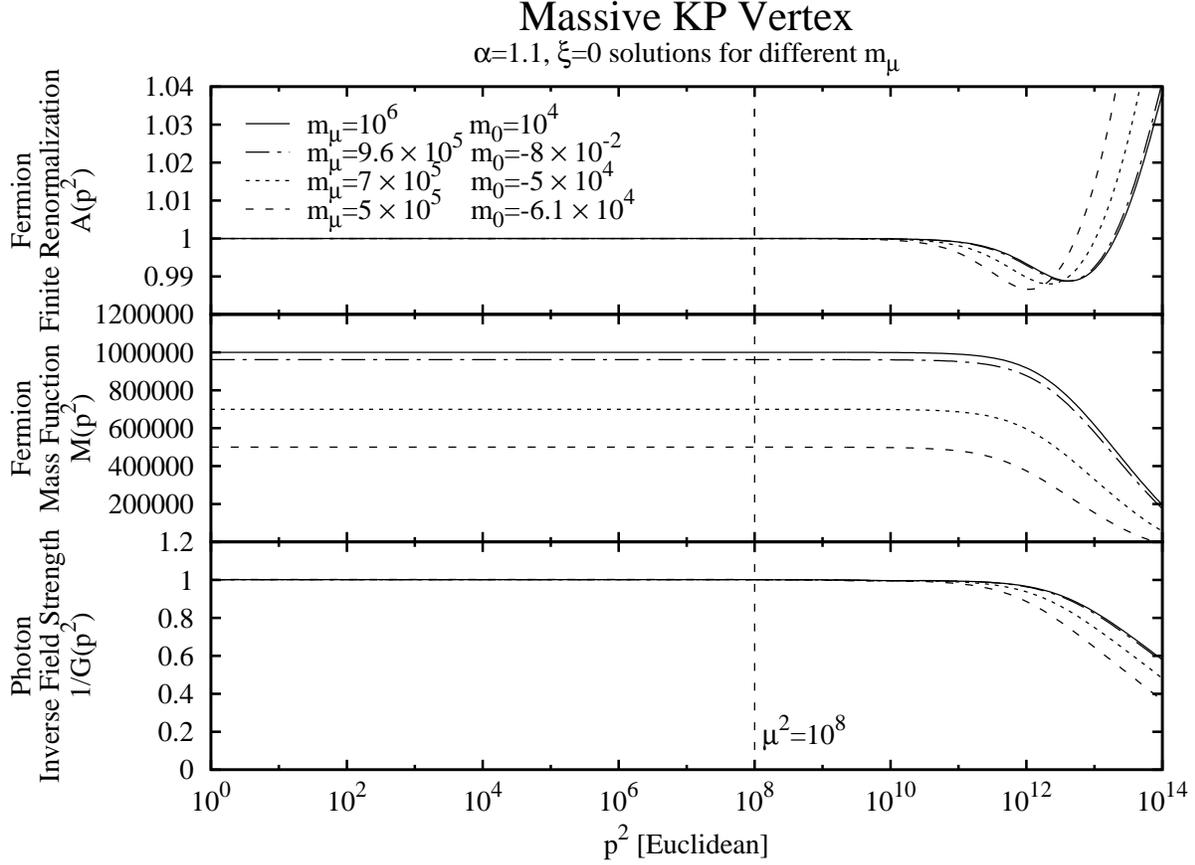}
\end{center}
\caption{High alpha high mass decreasing solutions $\alpha=1.1,\,\xi=0,\,m_\mu=10^6 \to 5 \times 10^5$.}
\label{fig:kp-sup-mass+chiral2-1}
\end{figure}

This implies a lower bound on the mass of high alpha solutions.
This lower bound can be seen clearly in Fig.~\ref{fig:kp-sup-mass+chiral2-1}, we lower the mass of $\alpha=1.1, \xi=0$, KP vertex solutions from $m_{\mu}=10^{6}$ until they failed to converge at $m_{\mu}=4\times10^{5}$, confirming the existence of a lower bound on the mass.

The solution with zero bare mass is also included in the figure because if the theory supports DCSB, this solution must exist, implying that the lower bound is lower than the chiral solution. However, it appears to be not much lower. By way of contrast, the quenched theory admits a solution for all masses below the chiral solution, but with oscillations in the mass function.
%
%
%
\section{\label{sec:Conclusion} Conclusion}

This paper studies Quantum Electrodynamics (QED) in 4-dimensions in the strong coupling region where the 
interactions between the fermions and photons are strong.  The Schwinger-Dyson Equations  
(SDEs) make it possible to analyse the field theory in this non-perturbative region since these equations are the field equations of that theory. The difficulty in  working with them arises from the fact that they are an infinite tower of non-linear integral equations and to solve these 
equations even for the 2-point Green's functions requires a meaningful truncation of this infinite system. Although such truncation is inevitable for 
solving these equations it must not alter the physics. Along this line the gauge invariance and the multiplicative renormalisability of the theory  
must be respected for every acceptable truncation scheme. During last five decades many studies have employed various truncations, such as Rainbow Ladder and others. Almost all of the analysis for these truncation schemes were  done using the quenched approximations, where the fermion loops are ignored and hence the photon propagator is treated as the bare one.  Although these studies have helped us understand how these equations behave and how to extract the physical quantities  they are not complete until we are able to study the full (dressed) theory.  

To date the only exploration beyond 
quenched theory was done by approximating the photon propagator to its first order perturbative expression\cite{Kondo:1990ig,Kondo:1990st}, which made it possible to study fermion and photon coupled system. This treatment serves as guide to understanding how this coupled system works in terms of its components however to determine the complete non-perturbative dynamics one needs to go beyond the quenched theory to the unquenched theory where we can analyse the strongly coupled fermions and photons. 

This goal is achievable now that a more realistic fermion-photon vertex\cite{Kizilersu:2009kg} has  become available. By making use of this vertex in this paper we have studied in depth unquenched QED in 4-dimensions by solving SDEs numerically for a fermion-photon propagator coupled system. 

This work deals with the renormalised unquenched SDE for Fermion wavefunction renormalisation, $F$, the mass function, $M$, and the photon wavefunction renormalisation, $G$, simultaneously for the vertices most commonly used (Bare, modified CP and BC ) in the past together with the new KP  vertex.  This is one of the very first and comprehensive study of the unquenched QED$_4$  which is compared against the quenched calculations and analysed for the vertices mentioned above to conclude which one of them perform better or worse based on the physics they must obey.  

We reported here that Curtis-Pennington vertex has a dynamical problem in the photon SDEs hence the solutions do not converge. For this reason we used the {\it{modified}} CP which includes CP vertex for fermion SDE and the BC for the photon SDE. The Bare, modified CP and BC vertices fail the {\it{gauge invariant photon wave-function renormalisation test}}, only the K{\i}z{\i}lers{\"u}-Pennington (KP) vertex leads to a highly gauge independent photon wave-function renormalisation. All the propagator functions must respect Multiplicative Renormalisability (MR) and a consequently the effective coupling and the mass function must be {\it{Renormalisation Group Invariant quantities}}.  While the Bare vertex fails to satisfy this criteria, the modified CP and KP vertices pass this test since they were both constructed to respect MR.  

We expect that the renormalised quantities are insensitive to UV cut-off, $\Lambda^2$, however BC vertex fails to display this  property for the fermion wave-function renormalisation, all other vertices realise this insensitivity to the cut-off. 

When the coupling was increased for all the vertices we saw that the photon wavefunction renormalisation experienced a limiting value for the coupling above which the $1/G$ has a zero crossing and therefore the solutions stop converging. We interpreted this phenomenon as a realisation of the
{\it{Landau Pole}} beyond which there are no solutions. To explore this phenomena we renormalised our propagator functions at the UV cut-off and, as was expected, we could raise the limiting value of coupling to one higher than to when the renormalisation point  was lower than the UV cut-off. For the massive solutions we saw that the Landau Pole can be avoided at momenta below the cut-off for high $\alpha$ solutions if the fermions have very large masses. We also showed that by unquenching the theory the tail of the solutions in the asymptotic region do not exhibit the power-law behaviour due to the broken scale invariance. It is interesting to speculate that if the cut-off is made large enough, then a Landau pole will always occur no matter how large the fermion mass is made or how small the coupling.

This study presents one of the first comprehensive analysis of various fermion-photon vertices and  their roles in SDE for the fermion and photon propagators.  We concluded that the K{\i}z{\i}lers{\"u}-Pennington vertex is superior to all other existing vertices for the full strong coupling QED$_4$. It shows the importance of having an appropriate unquenched vertex for the unquenched SDE studies by ensuring that the solutions satisfying the necessary criteria.

We will next examine Dynamical Mass Generation in QED$_4$ using the KP vertex, as well as  studying three dimensional QED as a toy model, 
since it presents Dynamical Chiral Symmetry Breaking as well as the confinement. Finally after these investigations we will turn our focus to QCD.
        
%
%

\begin{acknowledgments}
The authors are grateful to thank  Prof.A.W. Thomas for supporting this study and providing ideal working environments. 
We also acknowledge support from the Australian Research Council International Linkage Award (LX 0776452). This work supported by the Australian Research Council Discovery grant (DP0558878). We would like to 
thank Prof.M.R. Pennington and Prof. C.D. Roberts for useful discussions. We would like to extend our gratitude to eResearch SA  for providing us a platform to perform our numerical studies.
\end{acknowledgments}

%
%

\appendix
\section{\label{sec:first cycle}First Cycle Propagator Solutions of  SDEs}{\label{sect:first-cycle-propagators}}

\noindent
In this appendix we obtain the analytical first iteration {\it{cycle}} solutions of the fermion and photon propagator functions $F$, $M$ and $G$ which are stated in Eqs.(\ref{eq:submass},~\ref{eq:subphoton}).
In order to start the iteration process we first initialise these functions by choosing  $F(p^2)=1$, $G(q^2)=1$ and $M(p^2)=m_\mu$ in 
Eqs.(\ref{eq:submass},~\ref{eq:subphoton}).  The fermion wavefunction renormalisation, Eq.~(\ref{eq:submass}),  then reduces to~:
\be
F(p^2,\mu^2) &=& 1 +   {\overline{\Sigma}}_d (p^2,\mu^2) - {\overline{\Sigma}}_d (\mu^2,\mu^2)   \,,
\ee
\noindent
Inserting $F(p^2)=1$, $G(q^2)=1$ and $M(p^2)=m_\mu$ into the fermion self energy and its components, Eqs.~(\ref{eqn:dirac_self_energy},~\ref{eqn:sigmad_L},~\ref{eqn:sigmad_T}), we see that the total contribution comes from ${\cal{I}}^L_{\overline{\Sigma}_d}$ and nothing from ${\cal{I}}^T_{\overline{\Sigma}_d}$~:
\be
{\overline{\Sigma}}_d (p^2,\mu^2) & = & -\frac{\alpha}{4\,\pi^3}\, \int_E\, d^4k \frac{1}{p^2q^4}\,\frac{1}{(k^2+m^2)}\,
                                                                \Bigg\{ \xi \, \left[  p^2 k \cdot q +m^2 p \cdot q \right]+ \left[  2 \Delta^2 + 3q^2 k \cdot p \right]          \Bigg\}\,.  \nn\\
\label{eq:firstcyclesigmad}
\ee

\noindent
Performing the angular and radial integrals on Eq.(\ref{eq:firstcyclesigmad}) yields the first iteration cycle solution of the fermion self-energy 
\be
{\overline{\Sigma}}_d (p^2,\mu^2) & = & \frac{\alpha \xi}{4\,\pi}\, \Bigg \{ \ln\frac{p^2+m^2}{\Lambda^2+m^2} + \frac{m^2}{p^2} - \frac{m^4}{p^4} \ln\frac{p^2+m^2}{m^2}     \Bigg \} \,,
\label{eq:firstsigmad}
\ee
and the exact first cycle fermion wave-function renormalisation is~:
\be
F(p^2,\mu^2,m^2) & = & 1+ \,\frac{\alpha \xi}{4\,\pi}\, \Bigg\{   \ln \frac{p^2}{\mu^2} + \left[ \ln \left(1 +\frac{m^2}{p^2}\right) +   \frac{m^2}{p^2} - \frac{m^4}{p^4} \ln \left(1+\frac{p^2}{m^2} \right)  \right]   \nn  \\
       && \hspace{2.9cm}  -  \left[    \ln \left(1 +\frac{m^2}{\mu^2}\right) +   \frac{m^2}{\mu^2} - \frac{m^4}{\mu^4} \ln \left(1+\frac{\mu^2}{m^2}\right)       \right] \Bigg\}\,.
       \label{eq:F_firstcycle}
\ee
\noindent
Observe that in Equation (\ref{eq:F_firstcycle} ) all the momenta, mass and the renormalisation point 
appear as  ratios of $p^2$, $m^2$ and $\mu^2$. By inspection we can write
  \be
 F(p^2,\mu^2,m^2)  = F(p^2/s^2,\mu^2/s^2,m^2/s^2) \,.
        \label{eq:F_scaled}
 \ee

Going through the similar process for the mass function in Eq.(\ref{eq:submass})  we get
\be
M(p^2,\mu^2,m^2) &=& m + \left[ m  {\overline{\Sigma}}_d (p^2,\mu^2) + {\overline{\Sigma}}_s (p^2,\mu^2) \right] - \left[ m  {\overline{\Sigma}}_d (\mu^2,\mu^2) + {\overline{\Sigma}}_s (\mu^2,\mu^2) \right]\,,
\label{eq:firstcyclemass}
\ee
using the scalar part of the fermion self-energy, Eqs.(\ref{eqn:scalar_self_energy}, \ref{eq:sigmas_L}, \ref{eq:sigmas_T}) 
\be
{\overline{\Sigma}}_s (p^2,\mu^2,m^2) & = &   \frac{\alpha\, m}{4\,\pi^3}\, \int_E\, d^4k \,\frac{1}{q^2}\,\frac{1}{(k^2+m^2)}\, \Bigg\{
\frac{\xi}{q^2}\, \left[  k \cdot p  - p \cdot q  \right] +3
\Bigg\}\,,
\label{eq:firstcyclesigmas}
\ee
\noindent
once again we only have contributions from  ${\cal{I}}^L_{\overline{\Sigma}_s}$  to the first cycle calculations whereas ${\cal{I}}^T_{\overline{\Sigma}_s}$ does not contribute.
Integrating Eq.(\ref{eq:firstcyclesigmas}) yields the scalar part of the fermion self-energy 
\be
{\overline{\Sigma}}_s (p^2,\mu^2) & = & \frac{\alpha m}{4\,\pi}\, \left(\xi +3 \right)\, \Bigg \{ \ln \frac{\Lambda^2+m^2}{p^2+m^2} +1 - \frac{m^2}{p^2} \, \ln \frac{p^2 +m^2}{m^2}    \Bigg\}\,, 
\label{eq:firstsigmas}
\ee
and the first cycle mass function using Eqs.(\ref{eq:firstsigmad}, \ref{eq:firstsigmas}) is
\be
M(p^2,\mu^2,m^2) &=& m \, \Bigg\{ 1 + \frac{\alpha}{4 \pi} \left[ \xi\,\frac{m^2}{p^2} - \left(3 +\xi\,\frac{m^2}{p^2} \right) \, \left( 1+\frac{m^2}{p^2} \right) \, \ln \left( 1+ \frac{p^2}{m^2} \right)
\right]  \nn \\
&& \hspace{1.2cm} - \frac{\alpha}{4 \pi} \left[ \xi\,\frac{m^2}{\mu^2} - \left(3 +\xi\,\frac{m^2}{\mu^2} \right) \, \left( 1+\frac{m^2}{\mu^2} \right) \, \ln \left( 1+ \frac{\mu^2}{m^2} \right)
\right]
\Bigg\} \,.
\label{eq:M_scaled}
\ee
By inspection we see that
\be
M(p^2,\mu^2,m^2) =s \, \times\, M(p^2/s^2,\mu^2/s^2,m^2) \,.
\ee

Repeating the same procedure for the photon wavefunction renormalisation and vacuum self-energy, Eqs.(\ref{eq:subphoton}, \ref{eqn:photon_self_energy}, \ref{eqn:pi_L}, \ref{eqn:pi_T}) yields

\be
\frac{1}{G(q^2,\mu^2)} &=& 1 + \left[  {\overline{\Pi}}(q^2,\mu^2) -  {\overline{\Pi}}(\mu^2,\mu^2) \right]\,, 
\label{eq:firstcycleG}\\[3mm]
 {\overline{\Pi}}(q^2,\mu^2) & =& \frac{\alpha N_F}{3 \pi^3}\, \int_E\, d^4k \, \frac{1}{q^2} \frac{1}{(p^2+m^2)\,(k^2+m^2)} \, \Bigg\{
 2 k \cdot p - \frac{8}{q^2}\, (\Delta^2 + q^2 k \cdot p)
 \Bigg\}\,.
 \label{eq:firstcyclepi}
\ee
Integrating Eqs.~(\ref{eq:firstcyclepi}) yields the exact first cycle expression for the vacuum self energy~:
\be
 {\overline{\Pi}}(q^2,\mu^2) & =& \frac{\alpha\, N_F}{3\,\pi}\,\left\{ P_1 + P_2 + P_3 +P_4 + P_5 + P_6  \right\} \,,
 \label{eq:G_firstcycle}
 \ee
where
\be
P_1 & =& -\left(\frac{16}{3}\right) \, \frac{\Lambda^6}{q^6} + 2 \left( 1 - 8 \frac{m^2}{q^2}\right)\, \frac{\Lambda^4}{q^4} + 2 \left(  1 - 4 \frac{m^2}{q^2} - 8 \frac{m^4}{q^4} \right) \,\frac{\Lambda^2}{q^2} \,,  \nn  \\[3mm]
P_2 & = &  \sqrt{R_{\Lambda}}\,\,\,\frac{m^2}{q^2} \,\Bigg\{ \left( \frac{16}{3} \right)\,\frac{\Lambda^4}{q^4} 
              +\frac{2}{3}\,\left(  -1 + 16\, \frac{m^2}{q^2}\right)\,\frac{\Lambda^2}{q^2}
              -\frac{13}{6}+ \frac{26}{3}\,\frac{m^2}{q^2} + \frac{16}{3}\, \frac{m^4}{q^4}   \Bigg\} \,,  \nn  \\[3mm]
P_3 &= &   \ln\left[\frac{1}{ 2}\,\sqrt{R_{\Lambda} }\,\,\,+\frac{1}{ 2}\, \frac{\Lambda^2}{m^2} - \frac{q^2}{8 m^2} + \frac{1}{2}\right]\,,  \nn   \\
P_4 & = & - \frac{1}{4} \,\left(1+\frac{4 m^2}{q^2} \right)  \, \,\left[ -\frac{13}{6}  + \frac{26}{3} \, \frac{m^2}{q^2} +\frac{16}{3}\, \frac{m^4}{q^4} \right] \,, \nn  \\[3mm]
P_5 & =&  2\, \left(1- \frac{2m^2}{q^2}\right)\, \sqrt{\left( \frac{1}{4} + \frac{m^2}{q^2}\right)}\, 
                 \ln \frac{\frac{q^2}{m^2}\,\left(-\frac{\Lambda^2}{m^2}+\frac{q^2}{4 m^2} +1\right) + 2\, \sqrt{\frac{q^2}{m^2}\,\left(\frac{q^2}{4 m^2} +1\right)\, R_{\Lambda}}}
                 {\left(\frac{\Lambda^2}{m^2} + \frac{q^2}{4 m^2}+ 1 \right)} \,, \nn  \\[3mm]
P_6 & =& -  2\, \left(1- \frac{2m^2}{q^2}\right)\, \sqrt{\left( \frac{1}{4} + \frac{m^2}{q^2}\right)}\, 
                 \ln \left[ \frac{ q^2}{m^2} +  2\, \sqrt{\frac{q^2}{m^2}\,\left(\frac{q^2}{4 m^2} +1\right)} \right]    \,,
\label{eq:Ps}
\ee
\be
R_{\Lambda} = \frac{\Lambda^4}{m^4} +2\, \left(1 -\frac{q^2}{4\,m^2}  \right) \, \frac{\Lambda^2}{m^2} + \left( 1+\frac{q^2}{4\,m^2} \right)^2 \,.
\ee

\noindent
and making use of the above expressions one can form the first cycle photon wavefunction renormalization using Eq.(\ref{eq:firstcycleG}).

Several observation may be made here, firstly, the first iteration cycle expression for the photon propagator, Eqs.(\ref{eq:firstcycleG}, \ref{eq:G_firstcycle}), above which was derived from  
${\cal{I}}^L_{\overline{\Pi}}$, Eq.(\ref{eqn:pi_L}) and ${\cal{I}}^T_{\overline{\Pi}}$, Eq.(\ref{eqn:pi_T}), did not contribute. Furthermore all the quadratic and higher powers of $\Lambda$ in Eq.(\ref{eq:Ps}) cancel each other out and do not create any spurious infinities. Moreover it is important to note here that in order to obtain the correct value of the photon wavefunction renormalization we had to collect the terms in such a way that there was numerical cancellation between them and this required very high precision (i.e. 64 bit processing).  

Secondly, when $p^2$ is at the cut-off the behaviour of the $1/G$ is $1-{\rm function(mass)}$ where that function increases as the mass decreases for small masses and vice versa for large masses.   
 
Similar to the fermion wave function renormalisation and mass function all the $p^2$, $m^2$ and $\mu^2$ dependence in this equation are in the form of ratios again hence 
the scaling also applies the photon wavefunction renormalisation as~:

\be
G(p^2,\mu^2,m^2) = \,G(p^2/s^2,\mu^2/s^2,m^2/s^2) \,,
\ee

To obtain  the high mass solutions from the low mass ones one makes use of above scaling relations by relabelling them as~:  

\begin{eqnarray}
F(p^2,\mu^2,m_{\rm{high}}^2) &=& F(p^2/s^2,\mu^2/s^2,m_{\rm{low}}^2)\,,\\
M(p^2,\mu^2,m_{\rm{high}}^2) &=& s\,\times\,M (p^2/s^2,\mu^2/s^2,m_{\rm{low}}^2)\,,\\
G(p^2,\mu^2,m_{\rm{high}}^2) &=& G (p^2/s^2,\mu^2/s^2,m_{\rm{low}}^2)\,,
\end{eqnarray}
where $s^2=m_{\rm{high}}^2/m_{\rm{low}}^2$. However, this scaling procedure alters the renormalisation point. In order to get the solutions at the original renormalization point , $\hat{\mu}^2$, we can relate these scaled solutions to the desired ones using the first cycle analytic expressions from  Eqs.(\ref{eq:F_firstcycle},\ref{eq:M_scaled},\ref{eq:firstcycleG})~:

\be
F(p^2,\hat{\mu}^2,m_{\rm{high}}^2) &=&F(p^2,\mu^2,m_{\rm{high}}^2) - F(\hat{\mu}^2, \mu^2,m_{\rm{high}}^2)  +1 \\
M(p^2,\hat{\mu}^2,m_{\rm{high}}^2) &=& M(p^2,\mu^2,m_{\rm{high}}^2) - M(\hat{\mu}^2,\mu^2,m_{\rm{high}}^2)  +m_\mu \\
G(p^2,\hat{\mu}^2,m_{\rm{high}}^2) &=& G(p^2,\mu^2,m_{\rm{high}}^2) - G(\hat{\mu}^2,\mu^2,m_{\rm{high}}^2)  +1 
\ee

\noindent
In the region where $m^2 \ll \mu^2$ massless and massive solutions share the same UV tail on the other hand where  $ m^2 \gg \mu^2$  massless and massive solutions tails off the same IR  constant see 
Figs.~(\ref{fig:cmp4mass},~\ref{fig:kp-sup-mass+chiral2-1}). 

\bibliography{/Users/user/MYLIBRARY/myLibrary}
\end{document}